\documentclass[prb,amsmath,amssymb,showpacs,twocolumn]{revtex4}
\usepackage{graphicx}
\begin{document}
\title{Global Phase Diagram of the High Tc Cuprates}
\author{Han-Dong Chen}
\affiliation{Department of Applied Physics, McCullough Building,
Stanford University, Stanford CA 94305-4045}
\author{Sylvain Capponi}
\affiliation{Laboratoire de Physique Th\'eorique UMR 5152, Universit\'e
Paul Sabatier,  118 route de Narbonne, 31062 Toulouse, France}
\affiliation{Department of Physics, McCullough Building, Stanford
University, Stanford  CA~~94305-4045}
\author{Fabien Alet}
\affiliation{Theoretische Physik, ETH Z\"urich, CH-8093 Z\"urich,
Switzerland} \affiliation{Computational Laboratory, ETH Z\"urich,
CH-8092 Z\"urich, Switzerland}
\author{Shou-Cheng Zhang}
\affiliation{Department of Physics, McCullough Building, Stanford
University, Stanford  CA~~94305-4045}
\begin{abstract}
The high $T_c$ cuprates have a complex phase diagram with many
competing phases. We propose a bosonic effective quantum
Hamiltonian based on the projected $SO(5)$ model with extended
interactions, which can be derived from the microscopic models of
the cuprates. The global phase diagram of this model is obtained
using mean-field theory and the Quantum Monte Carlo simulation,
which is possible because of the absence of the minus sign
problem. We show that this single quantum model can account for
most salient features observed in the high $T_c$ cuprates, with
different families of the cuprates attributed to different traces
in the global phase diagram. Experimental consequences are
discussed and new theoretical predictions are presented.
\end{abstract}
\pacs{74.25.Dw, 71.30.+h,71.10.-w}
%74.25.Jb,74.25.Ha,
\maketitle

\def \cH{{\cal H }}
\def\nd{{^{\vphantom{\dagger}}}}
\def\yd{^\dagger}

\section{Introduction}
On the first look, the phase diagram of the high transition
temperature superconducting (HTSC) cuprates has a striking simplicity:
there are only three universal phases in the phase diagram of all
HTSC cuprates: the antiferromagnetic (AF), the superconducting
(SC) and the metallic phases, all with {\it homogeneous} charge
distributions. However, closer inspection shows a bewildering
complexity of other possible phases, which may or may not be
universally present in all HTSC cuprates. A large class of these
phases have inhomogeneous charge distributions. Because of this
complexity, formulating an universal theory of HTSC is a great
challenge. The $SO(5)$ theory unifies the AF and the SC order
parameters into a single five dimensional order parameter called
the superspin, and the effective quantum theory of the superspin
naturally explains proximity between the AF and the SC phases in
the observed phase diagram\cite{ZHANG1997}. The Goldstone modes of
the superspin fluctuations can be identified with the $\pi$
resonance mode observed in the neutron scattering
experiments\cite{DEMLER1995,DEMLER1998,ROSSATMIGNOD1991,
MOOK1993,FONG1995,DAI1996,FONG1996,MOOK1998,DAI1998,FONG2000,
FONG1999,HE2001,HE2002,STOCK2003,TRANQUADA2003}.
This theory also predicts the AF vortex
state\cite{ZHANG1997,AROVAS1997}, which has recently been observed
in a number of
experiments\cite{KATANO2000,MITROVIC2001,LAKE2001,LAKE2002,MILLER2002,
KHAYKOVICH2002,MITROVIC2003,KANG2003,FUJITA2003}.
Initially, the $SO(5)$ theory was motivated by the simplicity of
the pure AF and SC states, however, given the encouraging
agreements with the experiments, it is tempting to construct an
unified theory of the global phase diagram of the HTSC which
addresses the more complex inhomogeneous phases as well.
Complexities can of course be introduced phenomenologically into
the Landau-Ginzburg type of theories by simply introducing more
order parameters. However, this type of approach necessarily
limits the predictive power of theory. The goal of this paper is
to present a single effective quantum model of the superspin
degree of freedom, which can be derived systematically from the
microscopic electron models, and can be investigated reliably both
analytically and numerically. The global phase diagram of this
model is then compared with the experimentally observed phase
diagram of the HTSC cuprates.

When formulated on a coarse-grained lattice, with high energy
charge states projected out, the projected $SO(5)$ model describes
five local superspin degrees of freedom per
plaquette\cite{ZHANG1999}. These five states are the spin singlet
state at half-filling, the spin triplet states at half-filling,
and the singlet d-wave hole pair state. Using the Contractor
Renormalization Group (CORE) algorithm, Altman and
Auerbach\cite{ALTMAN2002A} showed that the projected $SO(5)$ model
can be systematically derived from the microscopic electron
models, and they also determined the parameters of the effective
$SO(5)$ model explicitly from the microscopic interaction
parameters (see also Ref.[\onlinecite{CAPPONI2002}]). Restricted
within the subspace of these five local states, the Hamiltonian
describing their propagation and interaction is completely
expressed in terms of bosonic operators and can be studied
reliably by the quantum Monte Carlo (QMC) calculations. The
simplest form of the projected $SO(5)$ model has been studied
extensively by the QMC method both in two dimensions
\cite{DORNEICH2002,RIERA2002,RIERA2002A} and in three dimensions
\cite{JOSTINGMEIER2002}. The overall topology of the phase
diagram, the scaling properties near the multi-critical point and
the nature of the collective excitations can be reliably obtained
from the QMC method, within the parameter regime of experimental
interests.

The simplest form of the quantum $SO(5)$ model describe either the
direct, first order transition from the AF to the SC state, or two
second order transitions with an uniform, intermediate AF/SC mix
phase in between the pure AF and the SC
states\cite{ZHANG1997,ZHANG1999}. In the case of the direct first
order transition as a function of the chemical potential, the
system at a fixed density is phase separated. However, in the HTSC
cuprates, there are other forms of charge and spin ordered states.
For example, neutron scattering cross section in $LSCO$ material
is peaked around $(\pi\pm\delta,\pi)$ and $(\pi,\pi\pm\delta)$,
where $\delta \sim 1/8$\cite{TRANQUADA1995,AEPPLI1997,WELLS1997}.
STM experiments have revealed periodic charge modulation with
period close to four lattice spacing\cite{HOFFMAN2002,HOWALD2003},
either near the vortex core or near surface impurities. In latter
case, alternative interpretation\cite{MCELROY2003,WANG2003,MCELROY2003A}
 based on quasi-particle interference is also possible, and
the two points of views are summarized by Kivelson {\it et al}.\cite{KIVELSON2003}.
 Motivated by these experiments, we extend the
simplest form of the projected $SO(5)$ model to include extended
interactions among the five bosonic states. In fact, these
extended interactions also arise naturally by carrying out the
CORE algorithm to extended ranges.

The projected $SO(5)$ model with extended interactions support a
more complex phase diagram. In particular, there are insulating
phases at fractional filling factors where the charges form a
lattice, usually commensurate with the underlying lattice. A
crucial aspect of this model is that all charge density wave 
states are formed by the Cooper pairs of the holes, rather than
the holes themselves\cite{CHEN2002A}. Throughout this paper, we
shall denote such states as the pair-density-wave (PDW) states or pair checkerboard states. This distinction has a profound experimental
consequence, since the real space periodicity of the former is
larger than the latter by a factor of $\sqrt{2}$. This type of
insulating PDW states is a consequence of strong pairing and low
superfluid density, a condition which is naturally fulfilled in
the underdoped cuprates, but has not yet been unambiguously
identified in other experimental systems before. The PDW state can
either take the form of stripes or checkerboards, depending on the
ratios of the extended interaction parameters in the model.
Furthermore, PDW states with longer periodicity generally requires
longer range interactions to stabilize. Based on this reasoning, a
simple picture emerges for the global phase diagram of underdoped
cuprates. The phase diagram consists of islands of insulating PDW
states, each with a preferred rational filling fraction, immersed
in the background of SC states (see Fig. \ref{FIG-types}). The
height of the Mott insulating PDW lobes vary depending on the
preferred filling fraction and the range of extended interactions,
but in principle, these insulating states are all self-similar to
each other, and similar to the parent AF insulator at
half-filling. There can be either a direct first order phase transition, or two second order phase transitions between the SC state and the PDW state, with the possibility of an intermediate ``supersolid" phase, where both orders are present.

Based on our model, the bewildering complexity of the cuprate
phase diagram can be deduced from a simple principle of the ``Law
of Corresponding States". This concept is borrowed from the work
of Kivelson, Lee and Zhang on the global phase diagram of the
quantum Hall effect\cite{KIVELSON1992}, in fact, our proposed
phase diagram in Fig. \ref{FIG-types} bears great similarity to
Fig. 1 of that reference. In the case of the QHE, the ``Law of
Corresponding States" physically relates {\it all} quantum phase
transitions at various filling fractions to a {\it single} quantum
phase transition from the $\nu=1$ integer state to the Hall
insulator. In recent years, this powerful mapping among the
different fractional states has been made more precise by the
derivation of the $SL(2,Z)$ discrete modular group transformation
from the Chern-Simons
theory\cite{PRYADKO1996,BURGESS2001,WITTEN2003}. Similarly, the
central idea of the current paper is to relate the fractional Mott
insulator to SC transition with the transition from the AF Mott
insulator at half-filling to SC state, which is already well
understood within the context of the original, simple $SO(5)$
theory. The construction of the Mott insulating states at various
fractional filling factors can be constructed from the ``Law of
Corresponding States", iterated {\it ad infinitum}, to give a
beautiful fractal structure of self-similar phases and phase
transitions, as presented in Fig. \ref{FIG-types}. The various
different compounds of the HTSC cuprates families have slightly
different microscopic parameters, and they correspond to different
slices of this global phase diagram. The global phase diagram
provides a basic road-map to understand the common elements and
differences among various HTSC compounds.

This paper mainly focuses on the zero temperature global phase
diagram of the underdoped cuprates. However, it is understood that
the model is valid below the pseudogap temperature, which we
interpret as the temperature below which the system can be
effectively described by the collective bosonic degrees of
freedom, like the magnons and the hole pairs. Therefore, it is
implied that the pseudogap state is a regime where the various
ground states discussed here compete with each other, and
different experiments may access different aspects of these
competing states. The existence of the pseudogap temperature gives
the fundamental experimental justification to investigate the
global phase diagram of the underdoped cuprates by a purely
bosonic model. In the future, we shall use the same model to
investigate the manifestations of these competing states at finite
temperature, in the pseudogap regime. A comparison of the charge
order predicted by this work and the STM experiment in the
pseudogap regime has recently been reported in Ref.[\onlinecite{chen2004}].

While this paper is presented within the logical context of the
$SO(5)$ theory, some of the ideas and results bear intellectual
similarities to the previous theoretical works. The idea of doped
holes forming ordered stripes has been discussed extensively in
Ref.
[\onlinecite{ZAANEN1989,TRANQUADA1995,TSUNETSUGU1995,KIVELSON1998,WHITE1998,ZAANEN1999}].
Although we focus more on the charged ordered states in the forms
of checkerboards of hole pairs, they are conceptually related to
stripes and can be realized experimentally or theoretically
depending on the microscopic parameters. The pseudogap temperature
was identified as the formation temperature of Cooper pairs by
Emery and Kivelson\cite{EMERY1995}. Our interpretation of the
pseudogap temperature is more general, which also includes the
formation of the magnetic collective modes in addition to the
holes pairs. Vojta and Sachdev\cite{VOJTA1999} have discussed the
phase diagram of doped Mott insulator with various charge ordered
insulating states at rational fractions. More recently, Zhang,
Demler and Sachdev have studied extensively the competition among
charge and spin order\cite{DEMLER2001,ZHANG2002}. Laughlin pointed
out that the small superfluid density in the underdoped regime is
responsible for various charge ordering
phenomena\cite{LAUGHLIN2002}. Haas {\it et al.}\cite{HAAS1995}
have noticed that the Wigner crystal state of the hole pairs could
be stabilized due to the competition of phase separation and long
ranged Coulomb interaction. Kim and Hor\cite{KIM2001} have
discussed experiments at certain ``magic" filling fractions in
terms of the commensurate Wigner crystal type of order of the
electrons, rather than the hole pairs discussed in this paper.
Restricted to the charge sector, the projected $SO(5)$ model is
essentially the same as the hard-core quantum boson model on a
lattice, whose phase diagram has been extensively
studied\cite{FISHER1989,PICH1998,HEBERT2002}.

This paper is organized as follows. In section
\ref{SECTION-model}, the projected $SO(5)$ model with extended
interactions is presented. The choice of parameters is discussed
from the CORE algorithm and phenomenology. The self-similarity of the
insulating states and the classification of the quantum phase transitions
are then discussed in section \ref{SECTION-heuristic}.
In section
\ref{SECTION-MF}, the global phase diagram of the model is
obtained within the mean field theory. The low energy collective
modes and their quantum symmetry is then studied using a
slave-boson approach. In section \ref{SECTION-QMC}, QMC simulation
is carried out to compare with mean-field results obtained in
section \ref{SECTION-MF}. The experimental consequences and
predictions are discussed in section \ref{SECTION-Exp}. Finally,
section \ref{SECTION-Conclusion} concludes our study.

\section{Hamiltonian of the model}\label{SECTION-model}

The effective bosonic $SO(5)$ model can be derived directly from
the microscopic Hubbard model or $t$-$J$ model, through a
renormalization group transformation called the Contractor
Renormalization(CORE) method\cite{MORNINGSTAR1996,ALTMAN2002A}. To
construct {\em bosonic} quasiparticles from fermionic model, we
divide the lattice into effective sites containing an {\em even}
number of sites. In order to conserve the symmetry between $x$ and
$y$-direction in the system, a plaquette of $2\!\times\!2$ sites
are typically chosen\cite{ZHANG1999,ALTMAN2002A}. In their CORE
study of the 2D Hubbard model, Altman and Auerbach
\cite{ALTMAN2002A} started from the spectrum of lowest-energy
eigenstates of the 2$\times$2 plaquette for 0, 1 and 2 holes,
respectively. The low energy eigenstates of the Heisenberg
plaquette can be determined easily. The nondegenerate ground state
$|\Omega\rangle$ (see Ref.[\onlinecite{ALTMAN2002A}] for a
real-space representation in terms of the microscopic states on a
plaquette) has energy $E_0\!=\!-2J$ and total spin $S\!=\!0$. This
``RVB" like singlet state will be the vacuum state of the
effective bosonic $SO(5)$ model. The next set of energy
eigenstates are three triplet states
$t^\dagger_\alpha|\Omega\rangle$ with energy $E_t\!=\!-J$ and
total spin quantum number $S\!=\!1$. All other energy eigenstates
of the Heisenberg plaquette have energies $E\!\ge\!0$ and can be
neglected in the low energy effective model. It should be noted
that the operator $t^\dagger_\alpha$ with spin $1$ and charge 0
create {\em hardcore bosons} because one cannot create more than
one of them simultaneously on a single plaquette. The ground state
of two holes is a ``Cooper''-like hole pair with internal d-wave
symmetry with respect to the vacuum.

\begin{figure}[t]
  \includegraphics{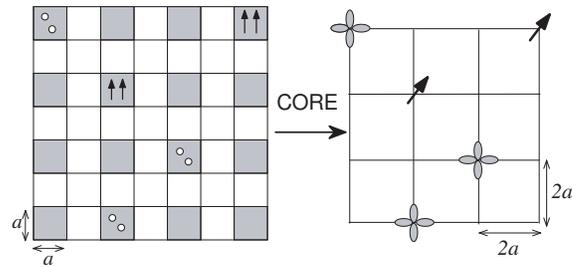}
  \caption{Illustration of the basic idea of the CORE method.
  To implement the CORE method, first decompose the original lattice
  in plaquettes, and then truncate the spectrum of a given plaquette to five lowest states,
  {\it i.e}, the singlet, hole-pair and three magnon states. An effective Hamiltonian for
  these bosons can then be determined using the CORE method.
  Left: local bosons in the original lattice. Gray rectangle denotes
  the singlet RVB vacua, circles denote holes and the set of two
  parallel vertical arrows denote the magnon.
  Right: local bosons on the lattice of plaquette. Leaf-like
  pattern denotes a local $d$-wave hole-pair on a plaquette.
  Canted arrow denotes local magnon on a plaquette.
  The singlet RVB vacuum is denoted by an empty site.}\label{FIG-CORE}
\end{figure}

Using the CORE method and keeping only the five lowest states (the
singlet boson, the three magnons $t^\dagger_\alpha$ and the
hole-pair $t^\dagger_h$), the effective Hamiltonian of these
bosons can be obtained as\cite{MORNINGSTAR1996,ALTMAN2002A,CAPPONI2002}
\begin{equation}
  \cH = \cH_0 + \cH_{ext},  \label{H0}
\end{equation}
where $\cH_0$ is the Hamiltonian of the previously studied $SO(5)$
model containing only on-site
interactions\cite{ZHANG1999,JOSTINGMEIER2002,DORNEICH2002}
\begin{subequations}
\begin{eqnarray}
    \cH_0 &=& \Delta_c \sum_i t^\dagger_h(i) t_h(i)
    +\Delta_s \sum_\alpha\sum_i t^\dagger_\alpha(i) t_\alpha(i)
    \nonumber\\
 &-&J_c \sum_{\langle ij\rangle} \left[ t^\dagger_h (i) t_h(j) + H.c. \right]
  \nonumber\\
 &-&J_s\sum_\alpha\sum_{\langle ij\rangle}
 \left[t^\dagger_\alpha (i) + t_\alpha(i)\right]
 \left[t^\dagger_\alpha (j) +t_\alpha(j)\right],
\end{eqnarray}
and $\cH_{ext}$ is the part containing extended interactions
\begin{eqnarray}
  \cH_{ext}&=&\left[V_c\sum_{\langle ij\rangle}
   + V_c' \sum_{\langle\langle ij\rangle\rangle}\right] n_h(i)n_h(j)
   \nonumber\\
  &-& J_\pi \sum_\alpha\sum_{\langle ij\rangle} \left[ t^\dagger_h (i)
t_h(j) t^\dagger_\alpha(j) t_\alpha(i) + H.c. \right] \nonumber\\
  &+& V_\pi\sum_\alpha
\sum_{\langle ij\rangle} [n_h(i) n_\alpha(j)+ n_h(j)n_\alpha(i)]
\nonumber\\ &+& \sum_{\langle ij\rangle}\sum_{S=0,1,2}
V_S\left(t_it_j\right)^\dag_S\left(t_it_j\right)_S.
\end{eqnarray}
\end{subequations}
The model is subjected to the hard-core constraint
\begin{eqnarray}
  \sum_\alpha n_\alpha(i)+n_h(i)\leq 1.  \label{hard-core}
\end{eqnarray}
Here, $\Delta_c$ and $\Delta_s$ are the energy costs to create a
hole-pair and magnon respectively. $J_c$ and $J_s$ are the hopping
terms of hole-pairs and magnons. $t_h(i)$ and $t^\dag_h(i)$ are
the annihilation and creation operators of hole-pair on plaquette
$i$. $t_\alpha(i)$ and $t^\dag_\alpha(i)$ are the annihilation and
creation operators of magnon on plaquette $i$ for $\alpha=x,y,z$.
$\langle\cdots\rangle$ and $\langle\langle\cdots\rangle\rangle$
denote nearest-neighbor(nn) and next-nearest-neighbor(nnn),
respectively. $n_h(i)=t^\dag_h(i)t_h(i)$ and
$n_\alpha(i)=t^\dag_\alpha(i)t_\alpha(i)$ are the hole-pair
density and magnon density operators on plaquette $i$,
respectively. The hole-pair density of $n_h$ per plaquette
corresponds to {\it twice} the real doping of $\delta$ holes per
lattice site, {\it i.e.}
\begin{eqnarray}
 \delta=n_h/2.
\end{eqnarray}
Finally, $\left(t_it_j\right)^\dag_S$ creates two magnons
simultaneously on plaquettes $i$ and $j$, which are coupled into
total spin $S$. ${\cal H}_{ext}$ contains nn and nnn hole pair
interactions $V_c$ and $V_c'$, exchange hopping terms $J_\pi$
between $t_\alpha$ and $t_h$ bosons, interaction $V_\pi$ between
nn magnons and hole-pairs. The 4-magnon interactions $V_{0,1,2}$
are important in the pure AF phase but we can neglect them since
we are mostly interested in the doped phase where magnon density
decreases.
According to the CORE calculation on 2-plaquettes and fixing
$J_s=1$ as the unit of energy, one obtains from the $t$-J model
with $J/t \simeq 0.4$ (relevant for cuprates)~:

\begin{table}[ht]
\begin{tabular}{|c|c|c|c|c|c|c|}
 \hline $J_c$& $J_{\pi}$ & $V_c$ & $V_0$ & $V_1$ & $V_2$ & $V_\pi$\\
\hline 2. &  -0.6 & 10 & -1 & -1 & 0.4 & -3 \\ \hline
\end{tabular}
\end{table}
\noindent For $J_c\sim 2$, ${\cal H}_0$ is approximately $SO(5)$
symmetric at the mean field
level\cite{ZHANG1999,JOSTINGMEIER2002,DORNEICH2002}.

The CORE derivation of the $SO(5)$ model (\ref{H0}) is only
approximate. It should also be born in mind that the $t-J$ and the
Hubbard models are also approximate models of the real cuprates
themselves. If one started from a different microscopic model (for
example, next-nearest neighbor hopping, extended Coulomb repulsion
and etc), one would have obtained a similar effective Hamiltonian
with different parameters. Therefore, in this paper, we shall take
the CORE parameters as a guidance, and study the robust properties
of the $SO(5)$ model with a more general set of parameters, as to
reproduce well-known results and compare directly with
experiments.

At half-filling ($n_h=0$), the model involves only the singlet and
the magnon, and the effective Hamiltonian containing only $J_s$
and $\Delta_s$ can be rewritten as
\begin{equation}
H= 2 J_s  \sum_{\langle ij\rangle}\frac{t^\dag_\alpha(i)+t_\alpha(i)}{\sqrt{2}}
\frac{t^\dag_\alpha(j)+t_\alpha(j)}{\sqrt{2}}  +
\frac{\Delta_s}{2} \sum_i L_{\alpha\beta}^2(i),
\end{equation}
where $[t^\dag_\alpha(i)+t_\alpha(i)]/\sqrt{2}$ is the AF moment
and $L_{\alpha\beta}(i)$ is the $SO(3)$ symmetry
generator\cite{ZHANG1999}
\begin{eqnarray}
  L_{\alpha\beta}(i)=  -i[t^\dag_\alpha(i)t_\beta(i)-t^\dag_\beta(i) t_\alpha(i)].
\end{eqnarray}

This model is similar to the nonlinear $\sigma$ model
 (NL$\sigma$M)\cite{CHAKRAVARTY1988,MANOUSAKIS1991}
\begin{equation}
 H= \rho_s \sum_{\langle ij\rangle}m_i^\alpha m_j^\alpha
 + \frac{1}{\chi} \sum_i S^2_i,
\end{equation}
where $m_i^\alpha$ is the $\alpha$ component of the AF moment and
$S_i$ is the angular moment on site $i$. After rescaling of time
using the spin velocity $c=\sqrt{\rho_s/\chi}$ and up to a
prefactor, the Lagrangian density of the NL$\sigma$M can be cast
in the usual form in the continuum limit
\begin{eqnarray}
\mathcal{L}_{NL\sigma}= \frac{1}{g} (\partial_x m)^2 + g
(\partial_\tau m)^2,
\end{eqnarray}
where $g\sim 1/\sqrt{\rho_s \chi}$ is a dimensionless constant.
This model has been studied extensively~\cite{MANOUSAKIS1991}. It
has a transition towards a disordered state at $g_c=1.45$. From
the computation of the staggered moment, we can find $g$ such that
the original Heisenberg value (0.3) of the AF moment is recovered:
$g_H=1.125$. On the other hand, we know from mean-field
calculations and QMC simulation that the disordered phase occurs
at $\Delta_s/J_s=8$. We then obtain the proportionality factor
between $g$ and $1/\sqrt{\rho_s \chi}$. Using $g_H$, we find that
an effective model for Heisenberg corresponds to
$\Delta_s/J_s=4.8$.

In most parts of this paper, we shall consider the simplified model
with $J_\pi=V_\pi=0$.

\section{Heuristic argument on the self-similarity between
pair-density-wave (PDW) states}\label{SECTION-heuristic} 
\begin{figure}[t]
  \includegraphics{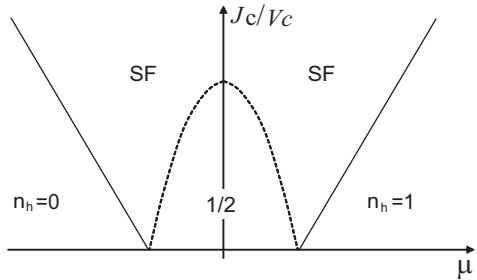}
  \caption{Phase diagram of a hard-core boson model with nearest-neighbor
   interaction. There are one superfluid (SF) state and three insulating states:
   zero-doping state ($n_h=0$), half-filled state ($n_h=1/2$) and fully-occupied
   state ($n_h=1$). At the next level of the hierarchy, longer ranged interactions
   lead to new insulating states with $n_h=1/4$ and $n_h=3/4$, as shown in
   Fig. \ref{FIG-hardcore-quartic}.}\label{FIG-hardcore-half}
\end{figure}

Let us ignore
for a moment the magnons, and consider a hard-core boson model
with extended interactions. The phase
diagram\cite{FISHER1989,PICH1998} of a hard-core boson model with
nn interaction contains one superfluid state and three
Mott-insulating states, corresponding to zero-doping ($n_h=0$),
half-filling ($n_h=1/2$) and fully-occupied ($n_h=1$), as sketched
in Fig. \ref{FIG-hardcore-half}. The half-filling state has
checkerboard charge order. If one transforms the hard-core boson
model into an AF Heisenberg model, the checkerboard order of the
bosons simply correspond to the AF order of the Heisenberg spins.
The following argument assumes that the checkerboard order at
half-filling is a basic and robust form of order, such that it is
repeated at all different levels of the hierarchy.

\begin{figure}[ht]
  \includegraphics{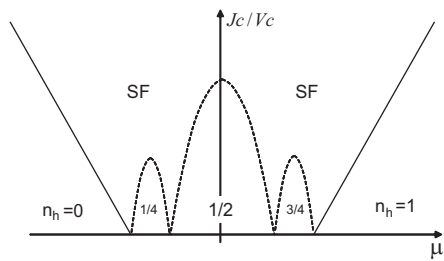}
\caption{Phase diagram of a hard-core boson model with nn and nnn
interactions. There are a superfluid (SF) state and five
insulating states with doping  $0/4, 1/4, 2/4, 3/4$ and $4/4$. At
the next level of the hierarchy, new insulating states are
developed at doping level $n_h=1/8$, $n_h=3/8$, $n_h=5/8$ and
$n_h=7/8$. This hierarchy construction can be iterated {\it ad
infinitum}, to obtain a self-similar phase diagram with insulating
phases at doping level $p/2^n$, with integers $p$ and $n$, such
that $0<p<2^n$. }\label{FIG-hardcore-quartic}
\end{figure}

%%%

If we regard the empty sites of the half-filled checkerboard state
as an inert background, we obtain a fully-occupied Mott-insulating
state on the coarse-grained lattice, with lattice spacing
$\sqrt{2}a \times \sqrt{2}a$. The nnn interaction on the original
lattice becomes the nn interaction on the coarse-grained lattice,
and a new half-filled checkerboard state can be stabilized on the
coarse-grained lattice. Such a state corresponds to doping
$n_h=1/4$ on the original lattice. Similarly, we can regard the
filled sites of the original half-filled checkerboard lattice as
an inert background, leaving with an empty state on the
coarse-grained lattice. A new checkerboard state can again be
stabilized on the coarse-grained lattice, which corresponds to
doping $n_h=3/4$ on the original lattice. This hierarchical
procedure of forming a new daughter checkerboard state from two
parent checkerboard state can obviously be iterated {\it ad
infinitum}, to obtain a fractal-like, self-similar phase diagram
as shown in Fig.\ref{FIG-hardcore-quartic}. It is interesting to
note that the nearest-neighbor interaction on the coarse-grained
lattice is just the next-nearest-neighbor interaction on the
original lattice. There is also the possibility that small regions
with coexisting SF and PDW orders (``supersolids") are present
around the Mott-insulating lobes in the phase
diagram~\cite{PICH1998,vanOtterlo1994,Batrouni1995,HEBERT2002}.

\begin{figure}[t]
    \includegraphics{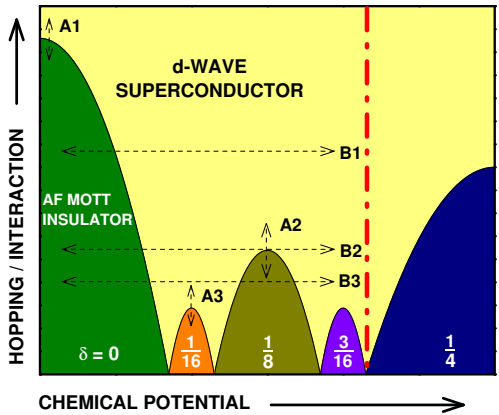}
\caption{A typical global phase diagram of the cuprates in the
parameter space of chemical potential and the ratio of boson
hopping energy over Coulomb interaction energy. This phase diagram
shows self-similarity among the insulating PDW states at
half-filling and other rational filling fractions. There are two
types of superfluid-insulator transition. The quantum phase
transition of ``class A" can be approached by varying the hopping
energy, for example, by applying a pressure and magnetic field at
constant doping. The quantum phase transition of ``class B" can be
realized by changing the chemical potential or doping. There can be either a direct first order phase transition, or two second order phase transitions between the SC state and the PDW state, with the possibility of an intermediate ``supersolid" phase, where both orders are present.
Different
families of cuprates correspond to different traces of ``class B".
For example, we believe $YBCO$ is $B1$-like, $BSCO$ may be close
to $B2$-like and $LSCO$ is $B3$-like. The vertical dash-dot line
denotes a boundary in the overdoped region beyond which our pure
bosonic model becomes less accurate. }
    \label{FIG-types}
\end{figure}

Having presented the generic phase diagram for the charge boson
only, we consider now the inclusion of the magnons in the full
$SO(5)$ with extended interactions. Generally, charge ordered
insulating states also have AF order. The $n_h=0$ state of the
charge boson corresponds to the undoped parent Mott insulator. The
$n_h=1/2$ state of the charge boson would correspond to
$\delta=1/4$ doping for the cuprates, which is probably at or
beyond the limit of applicability of our bosonic model. Therefore,
the phase diagram of the hard-core boson model in the range of
$0<n_h<1/2$ from Fig. \ref{FIG-hardcore-quartic} would translate
into a phase diagram of the cuprates in the doping range of
$0<\delta<1/4$, as shown in Fig. \ref{FIG-types}. As we shall show
later, this phase diagram is supported by accurate QMC
calculations of the $SO(5)$ model with extended interactions. We
expect that the insulating states of $\delta=1/16$ and $1/8$ are
AF ordered. Since the magnon density decreases with increasing
doping, the $\delta=3/16$ state may not be AF ordered.

The nature of the phase boundary between two different phases
shown in Fig. \ref{FIG-types} requires careful characterization.
We can classify all phase transitions into two broad classes.
Class $A$ describes transitions at fixed chemical potential,
typically at an effectively particle-hole symmetric point around
the tip of the Mott lobe. Class $B$ describes transitions where
the chemical potential or the density is varied. Each broad class
is further classified into three types, $1$, $2$ and $1.5$.
Generically, the phase transition between two ordered phase can be
either a single first-order transition or two second order
transitions, with a mixed state in between, where both order
parameters are nonzero. A third marginal possibility occurs at a
symmetric point, when these two second order phase transitions
collapse into a single one. In the context of high Tc cuprates,
these three types are shown in phase diagrams of
Ref.[\onlinecite{ZHANG1997}] as Fig. 1a,b,c, respectively. This
situation can be easily understood by describing the competition
in terms of a Landau-Ginzburg functional of two competing order
parameters\cite{Kosterlitz1976}, which is given by
\begin{eqnarray}
  F=\frac{1}{2}r_1\phi_1^2+\frac{1}{2}r_2\phi_2^2+u_1\phi_1^4+u_2\phi_2^4
  +2u_{12}\phi_1^2\phi_2^2
\end{eqnarray}
where $\phi_1$ and $\phi_2$ are vector order parameters with $N_1$ and $N_2$
 components, respectively. In the context of $SO(5)$ theory,
$N_1 = 2$ and $N_2 = 3$, and we can view $\phi_1^2$ as SC
component of the superspin vector, and $\phi_2^2$ as the AF
component of the superspin vector. These order parameters are
obtained by minimizing the free energy $F$. By tuning $r_1$, one
can drive a quantum phase transition from AF to SC. For
$u_{12}>\sqrt{u_1u_2}$, the quantum phase transition from AF to SC
is a single first order transition of ``type $1$". For
$u_{12}<\sqrt{u_1u_2}$, the transition from AF to SC consists in
two second order transitions, and there is a finite range of $r_1$
where AF and SC coexist uniformly; the transition is of ``type
$2$". For $u_{12}=\sqrt{u_1u_2}$, the phase transition occurs at
\begin{eqnarray}
    \frac{r_1}{\sqrt{u_1}}=\frac{r_2}{\sqrt{u_2}},
\end{eqnarray}
where the free energy takes the $SO(5)$ symmetric form
\begin{eqnarray}
  F=\frac{r_1\sqrt{u_1}}{2}\tilde{\phi}^2+u_{12}^2  \tilde{\phi}^4
\end{eqnarray}
with
\begin{eqnarray}
  \tilde{\phi}^2=\frac{\phi_1^2}{\sqrt{u_1}} + \frac{\phi_2^2}{\sqrt{u_2}}.
\end{eqnarray}
Since the free energy depends only on $\tilde{\phi}$,
 one order parameter can be smoothly rotated into the other
 without any energy cost. At this point, the chemical potential
is held fixed, but the SC order parameter and the charge density
can change continuously according the condition that
$\tilde{\phi}^2$ is constant. This is a special case intermediate between
``type $1$ and $2$" transitions, where two second order phase transitions
collapse into one. This transition can only occur at a $SO(5)$
symmetric point. We thus classify it as ``type $1.5$". The full
quantum $SO(5)$ symmetry can only be realized in the class $A$
transition of ``type $1.5$". On the other hand, the static, or
projected $SO(5)$ symmetry can be realized in class $B$
transitions of ``type $1.5$".

In HTSC cuprates, the charge gap at half-filling is very large, of
the order of $U \sim 6eV$ , it is not possible to induce the
``class A1" transition from the AF to the SC state by
conventional means. However, the charge gap in the fractional
insulating states is much smaller, of the order of $J_c$, and it
is possible to induce the ``class A2" or ``class A3" insulator to
superconductor transition by applying
pressure\cite{Arumugam2002,Sato2000} or by applying a magnetic
field\cite{HAWTHORN2003,SUN2003}.

As the chemical potential or the doping level is varied, a given
system, roughly corresponding to a fixed value of the quantum
parameter $J_c/V_c$ , traces out different one dimensional slices
in this phase diagram, with typical slices $B1$, $B2$ and $B3$
depicted in Fig. \ref{FIG-types}. The nature of the phase
transition $B1$ is similar to that of the superspin-flop
transition discussed in Ref.[\onlinecite{ZHANG1997}]. In this
case, the phase transition from the AF to the SC state can be
further classified into ``types 1, 1.5 and 2", with the last two
cases leading to a AF/SC mixed phase at the phase transition
boundary. For lower values of $J_c/V_c$ , the trace $B3$
encounters the $\delta = 1/8$ insulating phase. The key signature
of this type of phase transition is that the SC $T_c$ will display
a pronounced minimum as the doping variation traces through the
$\delta = 1/8$ insulating state. Meanwhile, the AF ordering
(possibly at a wave vector shifted from $(\pi,\pi)$) will show
reentrant behavior as doping is varied. The phase transition
around the fractional insulating phases can again be classified
into types ``1, 1.5 and 2", with possible AF/SC, AF/PDW, SC/PDW
and AF/PDW/SC mixed phases.

We believe that the AF to SC transition
in the $YBCO$, $BCCO$ and the $NCCO$ systems corresponds to a ``class
B1" transition. These systems only have a AF to SC transition,
which can be further classified into types ``1, 1.5 and 2", but they
do not encounter additional statically ordered fractional
insulating phases. On the other hand, the phase transition in the
$LSCO$ system, where $T_c$ displays a pronounced dip at $\delta=
1/8$, correspond to the ``class B3" transition.

\section{Mean-field phase diagram of the model}\label{SECTION-MF}

\subsection{Four-sublattice ansatz and mean-field phase diagram}

\begin{figure}[ht]
\begin{center}
  \includegraphics[scale=1]{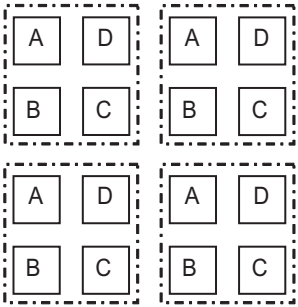}
  \end{center}
\caption{The schematic plot of the unit cell of a quarter-filled
magnetic insulating state. The solid square denotes a plaquette of
the original lattice and the dash-dot square denotes a
$2$-plaquette by $2$-plaquette unit cell.}
  \label{FIG-unitcell}
\end{figure}

Since the Hamiltonian (\ref{H0}) contains up to nnn interactions,
we can introduce the following four-sublattice ansatz within mean
field theory:
\begin{equation}
|\Psi\rangle = \prod_{r,m} \left[ e_m +h_m t_h^\dag(m,r)+x_m
t_x^\dag(m,r)\right]|\Omega\rangle
\end{equation}
where $|\Omega\rangle$ is the singlet ground state, $e_m, x_m,
h_m$ are real variational parameters, $m=A,B,C,D$ denote the sites
in a unit cell, and $r$ is the coordinate in the lattice of unit
cells, as sketched in Fig. \ref{FIG-unitcell}. The mean-field
energy $E_{MF}$ reads
\begin{widetext}
\begin{eqnarray}
 && \frac{4E_{MF}}{N} = \Delta_c\left(h_A^2+h_B^2+h_C^2+h_B^2\right)
  -2J_c\left[e_Ah_A+e_Ch_C\right]\left[e_Bh_B+e_Dh_D\right]
   +\Delta_s\left(x_A^2+x_B^2+x_C^2+x_B^2\right)
   \nonumber\\
   &&-4J_s \left[e_Ax_A+e_Cx_C\right]\left[e_Bx_B+e_Dx_D\right]
  -4J_\pi \left[h_Ax_A+h_Cx_C\right]\left[h_Bx_B+h_Dx_D\right]
  +2V_c [h_A^2+h_C^2][h_B^2+h_D^2] \nonumber\\
  &&+4V_c' [h_A^2h_C^2+h_B^2h_D^2]
  +2V_\pi\left([h_A^2+h_C^2][x_B^2+x_D^2]+[h_B^2+h_D^2][x_A^2+x_C^2]\right)
  \label{saddle-eng}
\end{eqnarray}
\end{widetext}
with the hard-core constraint
\begin{equation}
  h_m^2+x_m^2+e_m^2=1,\quad m=A,B,C,D.  \label{hardcore-MF}
\end{equation}
Here, $N/4$ is the number of unit-cells and $N$ is the number of
plaquettes.

\begin{figure}[ht]
    \includegraphics{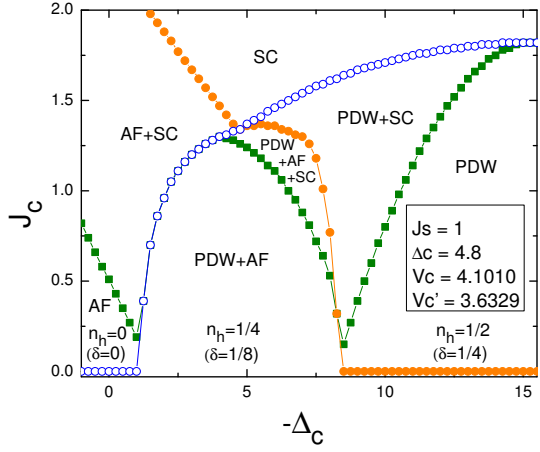}
\caption{The MF phase diagram obtained by minimizing the energy
functional (\ref{saddle-eng}) subjected to the hard-core
constraint (\ref{hardcore-MF}). $J_\pi$ and $V_\pi$ are taken to
be zero.}\label{FIG-phase-MF}
\end{figure}

By minimizing the energy functional (\ref{saddle-eng}) subjected
to the hard-core constraint (\ref{hardcore-MF}),
 we obtain the mean-field ground state
for a given set of parameters. Fig. \ref{FIG-phase-MF} plots the
mean-field global phase diagram, for $\Delta_s=4.8$, $V_c=4.1010$,
$V_c'=3.6329$ and $J_\pi=V_\pi=0$.

\begin{figure}[ht]
    \includegraphics{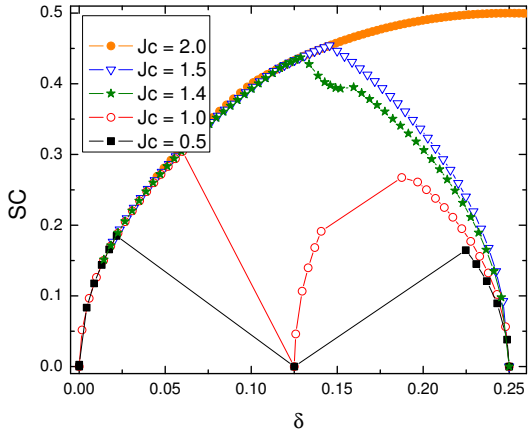}
\caption{Doping dependence of SC order parameter for different
$J_c$. For small $J_c$, there is a dip around hole-pair doping
$n_h=1/4$ (real doping $\delta=1/8$).}
    \label{FIG-SC-MF}
\end{figure}

\begin{figure}[ht]
   \includegraphics{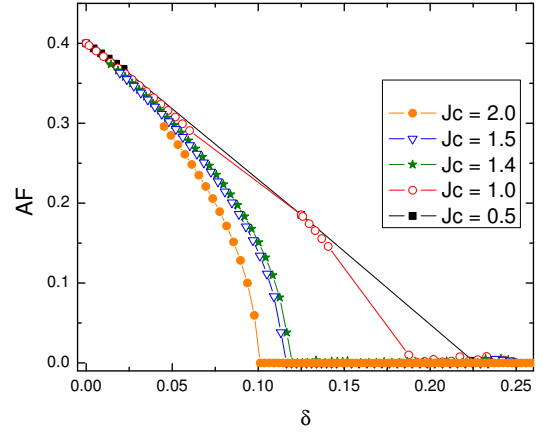}
\caption{Doping dependence of AF order parameter for different
$J_c$. For $J_c>1.3$, AF order decreases as doping increases and
vanishes around $\delta=0.1$.}
    \label{FIG-AF-MF}
\end{figure}

\begin{figure}[ht]
    \includegraphics{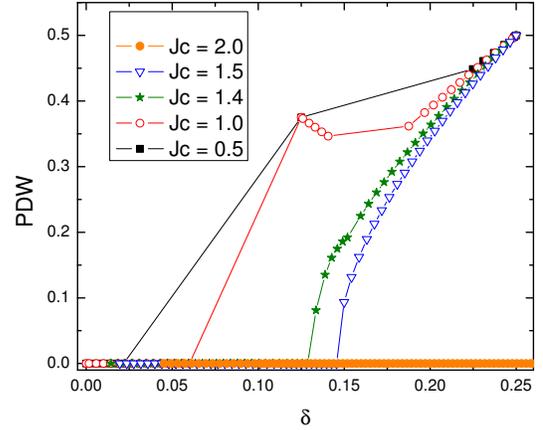}
\caption{Doping dependence of PDW order defined  for difference
$J_c$. For $J_c<1.3$, a strong peak is present at $n_h=1/4$ or
real doping $\delta=1/8$. }
    \label{FIG-CDW-MF}
\end{figure}

This phase diagram displays some rich features as expected. It has
three insulating state: an undoped antiferromagnetic (AF) state,
an insulating AF PDW state with hole-pair density $n_h=1/4$
($\delta=1/8$) and an insulating PDW state with hole-pair density
$n_h=1/2$ ($\delta=1/4$). Besides these insulating states, it also
has a pure SC phase, a supersolid phase and mix phases of
coexisting AF and SC order.

In Fig. \ref{FIG-SC-MF} and Fig. \ref{FIG-AF-MF}, we plot the
doping dependence of SC and AF orders for different $J_c$. If one
follows a ``class B1" trace, such as the one with fixed $J_c=1.5$,
the doping dependence of SC order mimics the behavior of $YBCO$
and $BSCO$ families with a underdoped region
$n_h<0.3(\delta<0.15)$ and an overdoped region $n_h>0.3
(\delta>0.15)$. If one follows a ``class B3" trace, such as the
one with fixed $J_c=1.0$, the SC order displays a pronounce dip
and the AF ordering is strongly enhanced around $n_h=1/4$
 ($\delta=1/8$). Therefore, the ``class B2" trace mimics the behavior
of $LSCO$ family.

The doping dependence of charge order parameter is also plotted in
Fig. \ref{FIG-CDW-MF}. It measures the charge modulation defined
by
\begin{equation}
  PDW = \frac{1}{4}\sum_{m}\left|n_h(m)-n_h\right|
\end{equation}
where $n_h$ is the average hole-pair density and $m$ is summed over
$A,B,C,D$. While ``class B1" trace shows no charge ordering in
underdoped region, ``class B3" trace displays a clear signature of
charge ordering around $\delta=1/8$.

\subsection{Slave-boson approach, effective Hamiltonian and
dynamical $SO(5)$ symmetry}

The hard-core constraint (\ref{hard-core}) can also be enforced by
introducing a slave-boson $t_e^\dag(i)$ for each lattice\cite{ZHANG1999}. The
presence of this boson indicates that the plaquette $i$ is empty.
The hard-core condition (\ref{hard-core}) is then replaced by
\begin{equation}
  \hat{Q}(i)=\left( \sum_\alpha t^\dag_\alpha t_\alpha(i)+
  t^\dag_ht_h(i)+t_e^\dag t_e(i)-1 \right)=0.\label{hard-core-slave-boson}
\end{equation}
This constraint can be enforced by introducing an additional site
dependent field $\lambda(i)$ and adding to the Hamiltonian
(\ref{H0}) an additional term
\begin{equation}
  H_{\lambda}= -\sum_i \lambda(i) Q(i).
\end{equation}
Since in physical states one always has one and only one boson per
lattice sites, destruction (creation) of a boson $t_a (t_a^\dag)$,
$a=x,y,z,h$, must always be accompanied by creation (destruction)
of the empty boson $t_e^\dag(t_e)$. In this way, the whole
Hamiltonian takes the form
\begin{eqnarray}
    \cH_{sb}&=&
      \Delta_c \sum_i t^\dag_h(i) t_h(i)
      +\Delta_s \sum_i t^\dag_\alpha(i) t_\alpha(i)
      \nonumber\\
      &&- J_c \sum_{\langle ij \rangle}
    \left[t^\dag_h(i)t_e(i)t_e^\dag(j) t_h(j)+H.c.\right]
             \nonumber\\ &&
     -J_s\sum_{\langle ij \rangle}
     \left[t^\dag_\alpha t_e(i)+t_e^\dag t_\alpha(i)\right]
    \left[t^\dag_\alpha t_e(j) +t_e^\dag t_\alpha(j)\right]
     \nonumber\\ &&
    -\sum_i \lambda(i) Q(i)
    +\cH_{ext}
%    - J_\pi \sum_\alpha\sum_{\langle ij\rangle}
%    \left[ t^\dagger_h (i)t_h(j) t^\dagger_\alpha(j) t_\alpha(i) + H.c. \right]
%      \nonumber\\ &&+
%    V_c\sum_{\langle ij \rangle}n_h(i)n_h(j)
%    +V_c'\sum_{\langle\langle ij \rangle\rangle}n_h(i)n_h(j)
%    + V_\pi\sum_\alpha \sum_{\langle ij\rangle}
%    [n_h(i) n_\alpha(j)+ n_h(j)n_\alpha(i)] .
    \label{H-sb}
\end{eqnarray}

By integrating out the $\lambda$ field in the partition function,
one automatically enforces the hard-core constraint $Q(i)=0$ on
each site. The saddle point approximation to (\ref{H-sb})
corresponds to replacing the boson operators and $\lambda$ field
with real constants to minimize the energy. Within the
four-sublattice ansatz, this is equivalent to minimizing the energy
functional subjected to the hard-core constraint
(\ref{hardcore-MF}). After obtaining the ground state, we can
expand the boson operators around their mean-field values and drop
quartic terms in the Hamiltonian (\ref{H-sb}) to yield an
effective Hamiltonian of the boson operators around the ground
state.

We shall now study the checkerboard state characterized by
following mean-field solution:
\begin{eqnarray}
    h_A=1, x_A=e_A=h_B=h_C=h_D=0
\end{eqnarray}
for the simplified model with $J_\pi=V_\pi=0$.

The saddle-point of the slave-boson Hamiltonian (\ref{H-sb}) can
be solved to yield
\begin{subequations}
\begin{eqnarray}
     x_D&=&x_B=x_C=0\\
     e_D&=&e_B=e_C=1\label{CB-solution-1}\\
     \lambda_D&=&=\lambda_B=\lambda_C=0\label{lambda_B1}\\
     \lambda_A&=& \Delta_c\label{CB-solution-2}
\end{eqnarray}
\end{subequations}
for $\Delta_s/Js\geq 4\sqrt{2}$ and
%\begin{subequations}
%\begin{eqnarray}
%     x_D&=&x_B=\sqrt{\frac{1}{16}\left[8-\frac{\Delta_s}{J_s}
%     \sqrt{\frac{64J_s^2+\Delta_s^2}{16J_s^2+\Delta_s^2}}~\right]}\\
%     %\quad\quad
%     x_C&=&\sqrt{\frac{1}{8}\left[4-\frac{\Delta_s}{J_s}
%     \sqrt{\frac{16J_s^2+\Delta_s^2}{64J_s^2+\Delta_s^2}}~\right]}\\
%     e_D&=&e_B=\sqrt{\frac{1}{16}\left[8+\frac{\Delta_s}{J_s}
%     \sqrt{\frac{64J_s^2+\Delta_s^2}{16J_s^2+\Delta_s^2}}~\right]}\\
%     %\quad\quad
%     e_C&=&\sqrt{\frac{1}{8}\left[4+\frac{\Delta_s}{J_s}
%     \sqrt{\frac{16J_s^2+\Delta_s^2}{64J_s^2+\Delta_s^2}}~\right]}\\
%     \lambda_D&=&\lambda_B=\frac{\Delta_s}{2}
%     -4J_s\sqrt{\frac{16J_s^2+\Delta_s^2}{64J_s^2+\Delta_s^2}}\label{lambda_B2}\\
%     %\quad\quad\quad\quad~~
%     \lambda_C&=&\frac{\Delta_s}{2}
%     -2J_s\sqrt{\frac{64J_s^2+\Delta_s^2}{16J_s^2+\Delta_s^2}}\\
%     \lambda_A &=& \Delta_c
%     \label{saddle-solution}
%\end{eqnarray}
%\end{subequations}
\begin{widetext}
\begin{subequations}
\begin{eqnarray}
     x_D&=&x_B=\sqrt{\frac{1}{16}\left[8-\frac{\Delta_s}{J_s}
     \sqrt{\frac{64J_s^2+\Delta_s^2}{16J_s^2+\Delta_s^2}}~\right]}
     \quad\quad
     e_D=e_B=\sqrt{\frac{1}{16}\left[8+\frac{\Delta_s}{J_s}
     \sqrt{\frac{64J_s^2+\Delta_s^2}{16J_s^2+\Delta_s^2}}~\right]}\\
     x_C&=&\sqrt{\frac{1}{8}\left[4-\frac{\Delta_s}{J_s}
     \sqrt{\frac{16J_s^2+\Delta_s^2}{64J_s^2+\Delta_s^2}}~\right]}
     \quad\quad\quad\quad\quad
     e_C=\sqrt{\frac{1}{8}\left[4+\frac{\Delta_s}{J_s}
     \sqrt{\frac{16J_s^2+\Delta_s^2}{64J_s^2+\Delta_s^2}}~\right]}\\
     \lambda_C&=&\frac{\Delta_s}{2}
     -2J_s\sqrt{\frac{64J_s^2+\Delta_s^2}{16J_s^2+\Delta_s^2}}
     \quad\quad\quad\quad\quad\quad\quad\quad
     \lambda_A = \Delta_c\\
     \lambda_D&=&\lambda_B=\frac{\Delta_s}{2}
     -4J_s\sqrt{\frac{16J_s^2+\Delta_s^2}{64J_s^2+\Delta_s^2}}\label{lambda_B2}
     %\quad\quad\quad\quad~~
     \label{saddle-solution}
\end{eqnarray}
\end{subequations}
\end{widetext}
for $\Delta_s/J_s\leq 4\sqrt{2}$.

The bosons are then expanded around their mean-field values as
\begin{subequations}
\begin{eqnarray}
    &&t_h(m,r)=h_m+\hat{b}_h(m,r),\label{expand-start}\\
    &&t_x(m,r)=x_m+\hat{b}_x(m,r),\\
    &&t_e(m,r)=e_m+\hat{b}_e(m,r),\\
    &&t_y(m,r)=\hat{b}_y(m,r),\\
    &&t_z(m,r)=\hat{b}_z(m,r).
    \label{expand-end}
\end{eqnarray}
\end{subequations}
Again, $m=A,B,C,D$ and $r$ is the coordinate in the coarse-grained
lattice of unit cells. Plug
(\ref{expand-start})-(\ref{expand-end}) into the slave-boson
Hamiltonian (\ref{H-sb}) and drop the quartic terms to yield the
quadratic effective Hamiltonian
\begin{eqnarray}
  H^{eff}=E_0+H_h^{eff}+H_x^{eff}+H_y^{eff}+H_z^{eff}\label{CB-H-eff}
\end{eqnarray}
with $E_0$ the mean-field ground state energy given by
\begin{subequations}
\begin{eqnarray}
  E_0&=& 2N(\Delta_c-\lambda_B)x_B^2+N(\Delta_c-\lambda_C)x_C^2\nonumber\\
    &&+4N x_C e_C x_B e_B\label{CB-H0}
\end{eqnarray}
and $H^{eff}_x$,$H^{eff}_h$,$H^{eff}_y$ and $H^{eff}_z$
\begin{widetext}
\begin{eqnarray}
  H^{eff}_x&=&
      \sum_{m=B,C,D}\left[\sum_q(\Delta_s-\lambda_m)\hat{b}^\dag_x\hat{b}_x(m,q)
      -\sum_q \lambda_m \hat{b}^\dag_e\hat{b}_e(m,q) \right]
      \nonumber\\ &&
     -J_s\sum_q\bigg[4e_B x_B
       \biggl(\hat{b}^\dag_x(C,q)\hat{b}_e(C,q)+H.c.\biggr)
       +B \leftrightarrow C
%    +4e_C x_C \biggl(\hat{b}^\dag_x(B,q)\hat{b}_e(B,q)+H.c.\biggr)
    \bigg]
    \nonumber\\ &&-2J_s\sum_q\bigg[
    e_C[\hat{b}^\dag_x(C,q)+\hat{b}_x(C,-q)]
    +x_C[\hat{b}^\dag_e(C,q)+\hat{b}_e(C,-q)]\bigg]\times
    \nonumber\\ &&\quad
    \bigg[\cos q_x
    \biggl(
    e_B[\hat{b}^\dag_x(B,-q)+\hat{b}_x(B,q)]
    +x_B[\hat{b}^\dag_e(B,-q)+\hat{b}_e(B,q)]\biggr)
    +\cos q_y\biggl(B\leftrightarrow D\biggr)
%    + \cos q_y  \biggl(
%    e_D[\hat{b}^\dag_x(D,-q)+\hat{b}_x(D,q)]
%    +x_D[\hat{b}^\dag_e(D,-q)+\hat{b}_e(D,q)]\biggr)
    \bigg]\label{H-x}
\end{eqnarray}
\begin{eqnarray}
  H_h^{eff}  &= &
  \sum_q\left(\Delta_c-\lambda_B+2V_c\right)
     \left[\hat{b}^\dag_h\hat{b}_h(B,q) + \hat{b}^\dag_h
      \hat{b}_h(D,q) \right]
     +\sum_q\left(\Delta_c-\lambda_C+4V_c'\right) \hat{b}^\dag_h
     \hat{b}_h(C,q)\nonumber\\
     &+&\sum_{q} (\Delta_s-\Delta_c)
      \hat{b}_x^\dag\hat{b}_x(A,q)
     -\lambda_A\sum_q \hat{b}^\dag_e\hat{b}_e(A,q)
     -J_s\sum_q 8e_B x_B~[\hat{b}^\dag_x(A,q)\hat{b}_e(A,q) +
     \hat{b}^\dag_e(A,q)\hat{b}_x(A,q) ]
     \nonumber\\
     &-&2J_ce_B\sum_q \biggl[ \cos q_y \biggl(
   \hat{b}_e(A,q)\hat{b}_h(B,-q)+\hat{b}^\dag_h(B,-q)\hat{b}^\dag_e(A,q)
    \biggr) +\cos q_x\biggl(B\leftrightarrow D\biggr)\biggr]
%       + \cos q_x \left[
%   \hat{b}_e(A,q)\hat{b}_h(D,-q)+\hat{b}^\dag_h(D,-q)\hat{b}^\dag_e(A,q)
%    \right]
    \nonumber\\
   &-&2J_ce_Be_C\sum_q\biggl[\cos q_x \biggl(
   \hat{b}^\dag_h(C,q)\hat{b}_h(B,q)+\hat{b}^\dag_h(B,q)\hat{b}_h(C,q)
    \biggr)+\cos q_y\biggl(B\leftrightarrow D\biggr)\biggr]
     \label{H-h}
\end{eqnarray}

\begin{eqnarray}
&&  H_\alpha^{eff}=
  \sum_{q}(\Delta_s-\lambda_B)
  \left(\hat{b}_\alpha^\dag\hat{b}_\alpha(B,q)
  +\hat{b}_\alpha^\dag\hat{b}_\alpha(D,q)\right)
  +\sum_{q}(\Delta_s-\lambda_C)
  \hat{b}_\alpha^\dag\hat{b}_\alpha(C,q)
  +\sum_{q}(\Delta_s-\Delta_c)
  \hat{b}_\alpha^\dag\hat{b}_\alpha(A,q)
  \nonumber\\
    &&-2J_s e_C e_B\sum_q
    \biggl[\hat{b}_\alpha^\dag(C,q)+\hat{b}_\alpha(C,-q)\biggr]
    \biggl[\cos q_y\biggl(\hat{b}_\alpha^\dag(D,-q)+\hat{b}_\alpha(D,q)\biggr)
    +\cos q_x  \biggl(D\leftrightarrow B\biggr)
    \bigg],\quad\alpha=y,z. \label{H-alpha}
\end{eqnarray}
\end{widetext}
\end{subequations}
Here, the operators $\hat{b}(m,q)$ are the Fourier transforms
of the bosonic operators $\hat{b}(m,r)$
\begin{eqnarray}
    \hat{b}(m,q)=\sum_r e^{iqr} \hat{b}(m,r), \quad   m=A,B,C,D
\end{eqnarray}
where the summation of $r$ is over the lattice of unit cells.

This effective Hamiltonian has four decoupled parts, $H_h^{eff},
H_x^{eff}, H_y^{eff}$ and $H_z^{eff}$.
%All of these four parts
%have two different kinds of terms, the ``chemical potential'' term
%and the hopping term. The chemical potential term itself has two
%sources of contribution. Due to the hard-core nature, one must destroy
%the existing boson to create a new one on cite $m$. This leads to an energy cost
%given by $-\lambda_m$. For example, to create a $y$-magnon on sublattice $B$, the
%annihilation of the existing $x$-magnon requires both the chemical potential
%term $\Delta_s/2$ and the loss of hopping energy because of the
%break-down of the AF bonds. This leads to exactly the formula of $\lambda_B$
%as given by (\ref{lambda_B1}) or (\ref{lambda_B2}).
%The other contribution is just the normal chemical
%potential term for the creation of the new particle.
%
%Now, let us turn to the hopping term. With the help of the existing
%background field such as $e_m, x_m$ and $h_m$, the AF or SC couplings
%leads to an effective hopping or exchange term of the bosonic fluctuation field.
%For example, the fluctuation of the
%$y$-magnon on sublattice $B$,$C$ and $D$ will induce the
%fluctuation of AF order. Therefore, by taking advantage of the
%existing singlets and AF coupling, the effective hopping energy of
%$y$-magnons among sublattice $B,C$ and $D$ will be $J_se_Ce_B$. On
%the other hand, the fluctuation of $y$-magnon on sublattice $A$
%will not induce AF order because the absence of the singlet. Thus,
%the hopping of $y$-magnon of sublattice $A$ can not take
%advantage of the AF coupling.
%
%Among the four decoupled parts,
Among these four decoupled parts, $H^{eff}_y$ and $H^{eff}_z$ are
just the Hamiltonian for the gapped magnetic mode on sublattice
$A$ and gapless magnetic Goldstone modes on sublattice $B,C$ and
$D$ due to the spontaneous symmetry breaking of SO(3) spin
symmetry. These two gapless Goldstone modes correspond to the
gapless uniform rotation of the AF ordering from $x$-direction to
$y$ or $z$ direction. $H^{eff}_x$ is the Hamiltonian of gapped
$x$-magnon mode on sublattice $B,C$ and $D$ due to the
condensation of $x$-magnons on these sublattices. The remaining
$H^{eff}_h$ is of great interest for it is the Hamiltonian of the
charge modes and $x$-magnon mode on sublattice $A$.

Under the insulating lobe of checkerboard state, the charge modes
are gapped. The charge gap $\omega_0$ is given by the solution to
the quartic equation of the form:
\begin{equation}
  \omega_0^4 + C_2 \omega_0^2 + C_3 =0,\label{gap-euqation}
\end{equation}
where $C_2$ and $C_4$ are quadratic functions of $J_c$.  This
equation is quartic because fields with momentum $q$ and $-q$ are
coupled and there are $t^\dag t^\dag$ terms in the effective Hamiltonian.

The condition $C_3=0$ gives the lobe-shaped second order phase
boundary where one charge mode becomes soft and a phase transition
from the PDW state to the SC state occurs. However, one must be
careful since it is possible that the system takes a first-order
transition before the charge mode softens. This indeed happens for
lower left part of the lobe, where the PDW state takes a first
order transition to a AF+SC mixed state. In the following
discussion, we assume the tip and some of the left part of the
lobe survive, as in the case plotted in Fig. \ref{FIG-phase-MF}.

\begin{figure}[t]
    \includegraphics{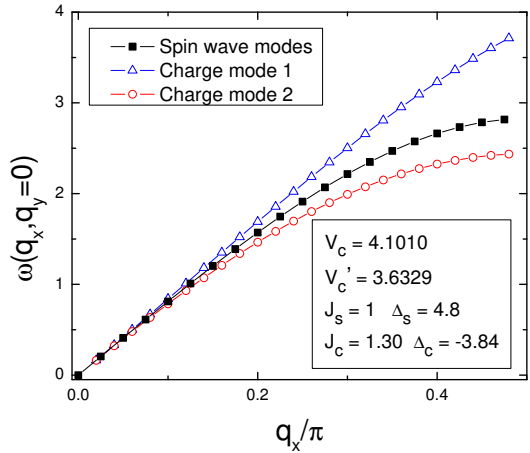}
\caption{Dispersion of four gapless modes at a quantum
multi-critical point, located tip of the $\delta=1/8$ checkerboard
lobe illustrated in Fig. \ref{FIG-phase-MF}. The two spin wave
modes have exactly the same dispersion due to the $SO(3)$ symmetry
of AF moment. For the parameters shown in the figure, these four
modes have the same speed at the multi-critical point. These four
modes correspond the four Goldstone modes required by the quantum
dynamical $SO(5)$ symmetry.}\label{FIG-modes}
\end{figure}

As one approaches the left part of the second order lobe following
a trajectory with constant $J_c$, the energy cost of removing
hole-pair decreases and becomes zero at the phase boundary. This
observation leads to the conclusion that a particle-like charge
mode becomes soft on the left part of the lobe. Similarly, one
can argue that a hole-like charge mode becomes soft at the right
part. At the tip, where the left part and right part of the lobe
meet, both particle-like and hole-like modes become soft.
Consequently, an effective particle-hole symmetry is dynamically
restored. One can check that $C_2$ indeed vanishes at the tip of
the lobe.

At the tip of the lobe, the speed of gapless charge modes is
determined by the interaction $V_c$ and $V_c'$. For appropriate
values of $V_c$ and $V_c'$, the speed of the gapless charge mode
can be the same as the speed of the two gapless AF spin wave
modes. When quantum $SO(5)$ symmetry is spontaneously broken to a
$SO(4)$ symmetry, there should be exactly four degenerate gapless
Goldstone bosons. This model shows that such a dynamical symmetry
is possible at the multi-critical point around the tip of the
$\delta=1/8$ lobe. Fig. \ref{FIG-modes} plots the dispersions of
four gapless modes at the tip of AF insulating checkerboard lobe
in the phase diagram of Fig. \ref{FIG-phase-MF}.

\section{Phase diagram obtained from Quantum Monte-Carlo
 simulation}\label{SECTION-QMC}

\subsection{Numerical simulation}

Because of the bosonic nature, the minus-sign problem is absent in
the quantum $SO(5)$ model. Therefore, simulations can be carried
out for systems with sizes much larger than the ones available
with fermionic QMC  simulation in the physically interesting
region. The pioneering numerical
works\cite{DORNEICH2002,JOSTINGMEIER2002,HU1999,HU2001,RIERA2002,RIERA2002A}
show that the projected $SO(5)$ model can give a realistic
description of the phase diagram of the HTSC cuprates and account
for many of their physical properties. In this section, we shall
present the simulation of the $SO(5)$ model with extended
interactions using the Stochastic Series Expansion (SSE)
method~\cite{SANDVIK1991,SANDVIK1997,SANDVIK1999A} with
operator-loop update~\cite{SANDVIK1999A}. This Quantum Monte Carlo
method was shown to be very efficient for simulations of harcore
bosonic systems~\cite{HEBERT2002,SCHMID2002,DORNEICH2002}. The
overall topology of the phase diagram agrees well with the mean
field calculation presented in the previous section, although the
parameters are strongly renormalized.

From the values given in section \ref{SECTION-model}, we see that
we can safely neglect $J_\pi$ which is rather small so that only
remains $J_c$ and $J_s$ which are both positive. In this section,
$J_\pi=0$.

In order to avoid the notorious sign problem in the Quantum Monte
Carlo simulations of the $SO(5)$ model with extended interactions,
all off-diagonal terms should be positive. On a square lattice
with only nearest-neighbour non-zero off-diagonal terms, the sign
of these matrix elements can be safely changed by a harmless
unitary transformation acting on hopping terms in only one of the
sublattices.

For each simulation, the number of loops (or ``worms'') made
during the loop operator update~\cite{SANDVIK1999A} is calculated
self consistently during the thermalization part, such that on
average the number of vertices visited by worms during each loop
operator update is equal to $C \langle n \rangle$. Here $\langle n
\rangle$ is the average number of non-Identity vertices in the
operator string (see Ref.~[\onlinecite{SANDVIK1999A}]) and $C$ a
proportionnality constant, usually taken between 1 and 5 (the
larger $C$, the more autocorrelations between successive Monte
Carlo configurations are reduced).

In order to plot the phase diagram, we should compute the order
parameters corresponding to AF, SC and PDW phase.We use the
superfluid density $\rho_c$ to locate SC phases. Indeed, $\rho_c$ can be related
to the winding numbers of the
world-lines which can be directly computed during the QMC
 simulations~\cite{POLLOCK1987,SANDVIK1997}.
 Here the winding
number only involves the charged particles, i.e. the hole pair
hopping. We can take a similar definition with the magnetic bosons
to define the spin stiffness.
%\begin{equation} \rho_s=\frac{1}{3}\frac{W^2}{2\beta}
%\end{equation}

It is straightforward to measure the density-density correlations and their Fourier
transform, the
structure factors $N(q_x,q_y)$ which indicate PDW phases~:
$$ N(\vec{q})=\frac{1}{L^2}\sum_{i,\vec{r}} exp(i \vec{q}.\vec{r}) \langle n_h(i) n_h(i+\vec{r}) \rangle $$
These quantities characterize the diagonal long-range order. On finite clusters,
the structure factor at the appropriate momentum diverges as the volume of the system in the
ordered phase, so that, by plotting  $N(q_x,q_y)/L^2$ vs $1/L$ ($L$ is the linear size of the system),
 a scaling analysis can demonstrate long range order.

Due to the intrinsic complexity of the projected SO(5) model and
to the large number of interaction types, we restricted the
simulations to small lattices ($4\times 4$ up to $10\times 10$
lattices) at low temperatures (typically $\beta=10$), to mainly be
in the ground state. Even using the powerful SSE method, we found
that for specific points in the phase diagram (near phase
boundaries for example), autocorrelation times of different
observables or tunneling times between two neighbouring phases can
be long, decreasing the statistical precision. This sometimes
prevents us from providing definitive statements about the nature
of phase transitions. However, outside of these regions, we can
clearly distinguish the nature of the different phases.

\subsection{Limiting cases}
The pure projected SO(5) model corresponds to $V_c=V_c'=V_\pi=0$
and $J_c=2 J_s$. This model has been studied with the same
technique~\cite{DORNEICH2002}. A first-order transition between AF
and SC phases was observed. It was already recognized that a small
$V_c$ and $V_c'$ were enough to turn this transition  into a
second-order phase transition.

Another interesting case occurs
when the triplet density becomes small. In that limit, the model reduces to one
type of hard-core boson with hopping $J_c$ and nearest and
next-nearest neighbor repulsion $V_c$ and $V_c'$ respectively.

\begin{eqnarray}
H &=& -J_c \sum_{\langle ij\rangle}\left[ t^\dagger_h (i) t_h(j) +
H.c. \right] \nonumber\\ &&+ \left[ V_c\sum_{\langle ij\rangle} +
V_c' \sum_{\langle\langle ij\rangle\rangle}\right] n_h(i)n_h(j)
\end{eqnarray}
so that we can fix $J_c=1$ as the energy unit. This model has been
extensively studied in Ref.~[\onlinecite{HEBERT2002}] and some results have
been obtained with the same SSE method. Let us review a few useful results.

\subsubsection{Half-filling}
The phase diagram at half-filling is well known and exhibits solid
phases with either $(\pi,0)$ (stripe) or $(\pi,\pi)$
(checkerboard) structures. In between exists a superfluid phase
with a non-zero superfluid density $\rho_c$. We recover the same results as Hebert {\it et al.}~\cite{HEBERT2002}~:
with our choice of interactions, by varying $J_c$, we can drive the system from a
superfluid toward a $(\pi,0)$ (stripe) phase.
%\begin{figure}
%\includegraphics{hebert.eps}
%\caption{Phase diagram at half-filling: QMC data, mean-field
%results (dotted line). Our model corresponds to the dashed line:
%it has a transition from a SC to a striped phase for $J_c\simeq
%1.3$. }
%\end{figure}

\subsubsection{Away from half-filling}
Away from half-filling, our grand-canonical algorithm is able to
check whether we are in a phase separated state or not by looking
at the histogram of the density during the
simulation. On general grounds, the presence of a double peak structure shows the
presence of true phase separation in the system.

%As a first case, we can look at the pure $V_c$ model, very close
%to half-filling. We clearly see a transition from insulating phase
%with $(\pi,\pi)$ order (checkerboard) to a superfluid phase. We
%also notice a jump in the density which is a sign of phase
%separation.

With our choice of parameters, we find that
away from the
striped state, there is a phase, close to half-filling, where both
$N(0,\pi)/L^2$ (PDW order parameter) and $\rho_c$ are finite, that is a supersolid phase.
Moreover, there is no sign of phase separation so that it is a
true homogeneous phase, as claimed in Ref.~[\onlinecite{HEBERT2002}].

\subsection{Global phase diagram}
\begin{figure}
  \includegraphics{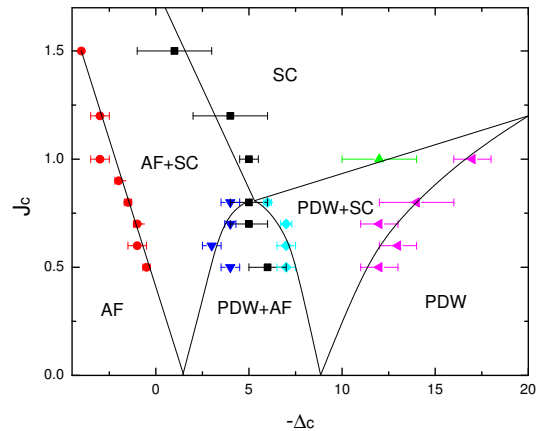}
\caption{The global phase diagram obtained by QMC. The parameters
used in simulation are $\Delta_s=4.8$, $V_c=4.1010$, $V_c'=3.6329$
and $J_\pi=V_\pi=0$. The lines are guides to the eye only. The
overall topology of the phase diagram agrees well with the
mean-field phase diagram of Fig. \ref{FIG-phase-MF}, although the
parameters are renormalized.} \label{FIG-Phase-QMC}
\end{figure}

\begin{figure}
\includegraphics{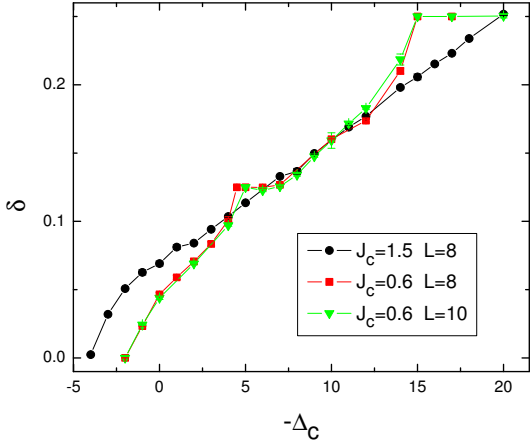}
\caption{\label{fig:delta} $\delta=n_h/2$ vs chemical potential
for $J_c=1.5$ and $J_c=0.6$. The presence of plateaux indicates
incompressible PDW phases. The finite size effects are rather
small.}
\end{figure}

Now that the parameters are fixed, we can compute the phase
diagram for various $J_c$ and chemical potential $\Delta_c$ (see
Fig.~\ref{FIG-Phase-QMC}). In order to discuss these results, we
plot in the following most of the data as a function of doping
$\delta=n_h/2$ which depends on the chemical potential $\Delta_c$
as shown on Fig.~\ref{fig:delta}, but has no strong finite-size
effects. Let us comment on some results.

\begin{figure}
\includegraphics{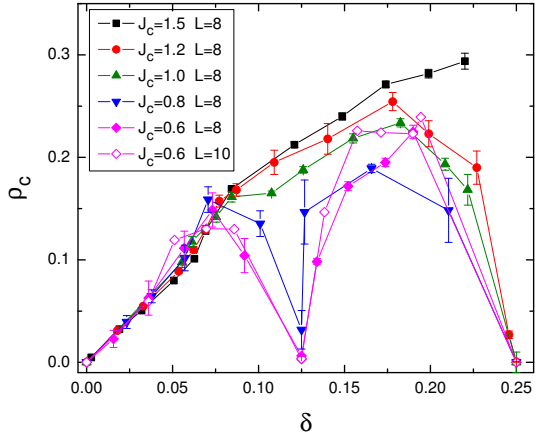}
\caption{\label{fig:SC} Superfluid density as a function of doping
for various $J_c$. The ``class B1" trace with $J_c=1.0$ bears
great similarity with the famous dome-like phase diagram of many
cuprate families. Also, the ``class B3"-like trace with
$J_c\leq0.8$ has a pronounced dip at $\delta=1/8$ as the phase
diagram of $LSCO$ family does.}
\end{figure}

\subsubsection{Large $J_c$}
\begin{figure}
\includegraphics{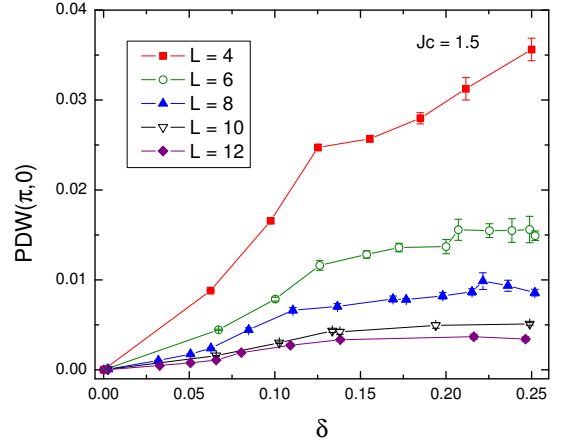}
\caption{\label{fig:CDW_J15} Scaling of $(\pi,0)$ PDW order
parameter ($N(\pi,0)/L^2$) for
various sizes ($L=6$ to 12 from top to bottom) with $J_c=1.5$.
We checked that it extrapolates to 0 for all $\delta$.}
\end{figure}

For all values of $J_c$, the superfluid density increases linearly
with doping, for small doping, thus capturing a key piece of the
Mott physics. For large $J_c$ ($J_c=1.5$ for example), reducing
hole density starting from $\delta=1/4$, we have a smooth decrease
of superfluid density (see Fig.~\ref{fig:SC}). At the same time,
magnon density increases and gives rise to AF.
 Fig.~\ref{fig:CDW_J15} shows typical data of $N(\pi,0)/L^2$ which is
the PDW order parameter for various
sizes. In order to get information on the thermodynamic limit, we
have performed a finite-size scaling of our data, and it vanishes for all
fillings. We have a superconducting state for all dopings, with a possible
coexistence with AF at low-doping. When the doping vanishes, we recover
a pure AF phase.

\subsubsection{Small $J_c$}
For smaller $J_c$, the long-range interactions between hole pairs start
to play a role.

\begin{figure}
\includegraphics{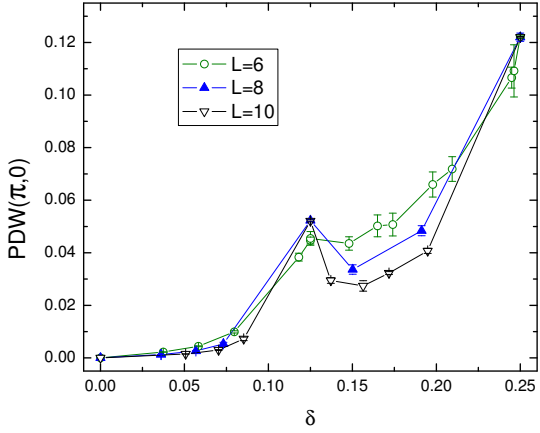}
\caption{\label{fig:CDWvsL} $(\pi,0)$ PDW order as a
function of $\delta$ for various sizes ($L=6, 8$ and $10$) for $Jc=0.6$.
We checked that, within the system sizes simulated,   finite PDW order  survives
in the thermodynamic limit for $n_h\geq 0.25$ ($\delta\geq0.125$).}
\end{figure}

{\it Large doping ---} The first example is the appearance of a
$(\pi,0)$ PDW insulating phase at $n_h=1/2$ ($\delta=1/4$). Indeed,
close to this doping, the magnon density is very small so that the
model is similar to the single hardcore boson case. We therefore
recover a transition at fixed $n_h=1/2$ ($\delta=1/4$) as a
function of $J_c$ between a superfluid phase (SC)
 and an insulating stripe phase (with a finite PDW order and a
 vanishing superfluid density).
 A finite-size scaling of the PDW order parameter showing a finite value is
shown on Fig.~\ref{fig:CDWvsL} for $n_h=0.5$ ($\delta=0.25$) and
$J_c=0.6$. On Fig.~\ref{fig:SC}, we see that below $J_c\sim 1.3$,
the superfluid density vanishes at $n_h=1/2$ ($\delta=1/4$) in
agreement with what has been found for the single-boson
model~\cite{HEBERT2002}. Mean-field results find a higher value
($V_c'/2\simeq 1.8$).

\begin{figure}
\includegraphics{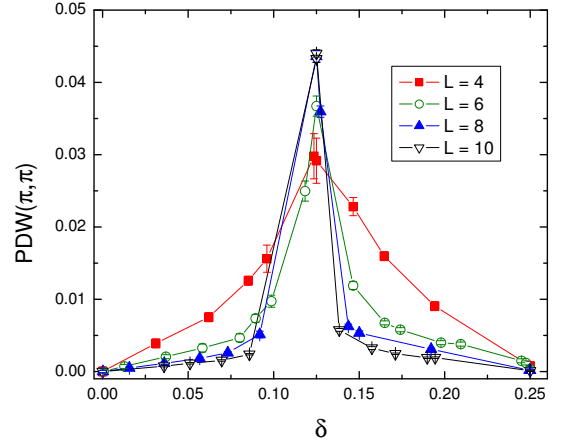}
\caption{\label{fig:CDW_pi_J07} $(\pi,\pi)$ PDW order as a
function of $\delta$ for various sizes ($L=4$, 6, 8 and 10) and
$J_c=0.7$. We checked that it is finite only for $n_h=1/4$ ($\delta=1/8$)
checkerboard.}
\end{figure}

{\it Intermediate doping ---} A second  example of the interaction
effect is given by the appearance of an insulating PDW phase at
$n_h=0.25$ ($\delta=0.125$) for low enough $J_c$ ($J_c\leq 0.7$).
A finite PDW order parameter and $\rho_c=0$ are shown again on
Fig.~\ref{fig:CDWvsL} and~\ref{fig:SC} for $J_c=0.6$. This state
corresponds to a localization of one hole-pair every 4 sites, so
that this  $n_h=1/4$ ($\delta=1/8$) checkerboard also possesses a
finite $N(\pi,\pi)/L^2$ as shown on Fig.~\ref{fig:CDW_pi_J07}. We
find that this $n_h=1/4$ ($\delta=1/8$) checkerboard is
insulating, with a vanishing  superfluid density. However, it
could be possible to have a supersolid phase\cite{Troyer}.

For intermediate dopings $0.25<n_h<0.5 (0.125<\delta<0.25)$ and
small $J_c$, we find a finite superfluid density and possibly a
finite PDW order, that is a supersolid region. We checked that this phase is
stable against phase separation. We cannot be conclusive about
 its extension in the phase diagram, due to the long
autocorrelation time for some observables. This explains the large
error bars in some parts of the phase diagram. However, since we
know that such a phase exists and is stable close to half-filling,
we can assume that it has a finite extension. It would be
interesting to check whether this supersolid also exists for a
single hard-core boson model close to $n_h=1/4$ ($\delta=1/8$), which
is much easier to simulate. It is remarkable that for $J_c$ values
close to the transition, we see a dip in the superfluid density
(Fig.~\ref{fig:SC}), which is due to the proximity to the
 insulating state.

\begin{figure}[ht]
  \includegraphics{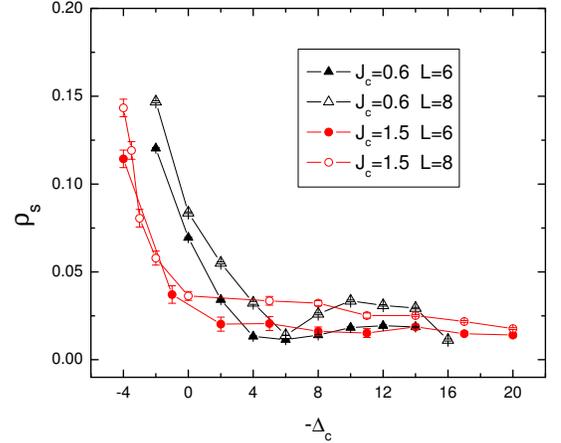}
  \caption{The spin stiffness versus chemical potential for different sizes and
  $J_c$.}\label{fig:AF}
\end{figure}

\subsubsection{Magnetic properties}
As shown on Fig.~\ref{fig:AF}, the spin stiffness shows
a monotonic decrease with chemical potential (or doping) so that
it is difficult to locate precisely where it vanishes.
For small doping $\delta$ or chemical potential $-\Delta_c$,
we observe a rapid linear
decrease so that we can estimate roughly where AF could
vanish in the thermodynamic limit. These phase boundaries are
shown on Fig.~\ref{FIG-Phase-QMC} and are in agreement with what has been found
at the mean-field level.

For small $J_c$, AF
seems to vanish at $n_h=1/4$ ($\delta=1/8$). However, our data of
$\rho_s$ also show a shallow peak above this filling, which could indicate a
reentrance of AF as found in mean-field. Unfortunately, with our
current limitation on available sizes and statistics, we cannot
conclude for sure about this possibility.

\subsection{Summary}
The qualitative features found at the mean-field level are still
present in a full numerical calculation and we have a very nice
overall agreement (see Fig.~\ref{FIG-phase-MF} and~\ref{FIG-Phase-QMC}).
Of course, exact critical values of $J_c$ for
superconducting-insulator transition are different from mean-field
values but this does not change the physical conclusions.

It is very difficult to point precisely where AF vanishes since
spin stiffness does not show any sharp drop. However, we clearly
see a strong reduction of $\rho_s$, which seems to decrease
linearly with doping. With the chosen parameters, it seems
plausible that both this line and the PDW transition line merge
close to the tip of the $n=1/4$ lobe as was found at the
mean-field level.  This result was associated to the dynamical
restoration of $SO(5)$ symmetry at this point. Our data might be
an indication that this is indeed the case  (we have taken the
same $V_c$ and $V_c'$ so it seems pretty robust). However, a
complete answer can only be provided by computing dynamical
correlations, which is more involved and require good statistics.

\section{Experimental consequence and predictions}\label{SECTION-Exp}
In this work we have constructed a single quantum Hamiltonian
based on the projected $SO(5)$ model with extended interactions,
and presented detailed analytical and numerical calculations which
give a consistent global phase diagram as depicted in Fig.
\ref{FIG-types}. This schematic phase diagram is obtained from the
quantitative model calculations summarized in Fig.
\ref{FIG-phase-MF} and \ref{FIG-Phase-QMC}. This model captures
the overall topology of the cuprates phase diagram, including the
dome-like feature of $T_c$, which is determined within our model
by the superfluid density, the $1/8$ anomaly due to charge
ordering, the coexistence of SC with AF and possibly with charge
order. As mentioned in the introduction, some of these
features have been discussed in other theoretical contexts before,
however, it is rather remarkable that they are all reproduced by a
single quantum model accessible by reliable QMC. Below we shall
discuss some of these features and present more detailed
theoretical predictions.

\subsection{Dependence of superfluid density on doping}
A remarkable feature of the HTSC cuprates is the dome-like
dependence of $T_c$ on doping. Experiments have also shown a
remarkable dependence of superfluid density on doping. On the
underdoped side, both $T_c$ and the superfluid density scales
linearly with doping, a fact commonly referred to as the ``Uemura
plot". Further examination also shows that the superfluid density
in the overdoped regime decreases with increasing doping, which is
commonly referred to as the ``boomerang effect". In cuprate
system, the muon spin relaxation rate $\sigma(T\rightarrow 0)$ is
proportional to $n_s/m^*$ where $n_s$ is the superfluid density
and $m^*$ is the effective mass of hole-pairs\cite{UEMURA2001}. To
explain the deviation from the linear relation between $\sigma$
and doping in the overdoped regime, Uemura proposed that some of
the doped holes do not form pairs and are phase separated from the
SC hole-pairs, even at zero temperature\cite{UEMURA2001}. A
similar phase separation picture was proposed by
Uchida\cite{UCHIDA2003}. However, it has always been rather
puzzling that if the phase separated normal electrons existed at
zero temperature, they should provide a channel of relaxations,
which has not been observed experimentally.

Within our effective bosonic model, $T_c$ is directly determined
by the superfluid density $\rho_c$. As we can see from Fig.
\ref{fig:SC}, $\rho_c$ indeed scales linearly with the doping
density for small doping, and has a dome-like dependence for
higher doping for intermediate values of $J_c$. It peaks around
$\delta\sim 18\%$ for $J_c=1.2$. The physical reason for the
behavior on the overdoped side arises from the tendency of hole
pairs to form a competing charge ordering state at either $3/16$
or $1/4$. Within this picture, when more holes are added into the
system on the overdoped side, the holes are still paired, but some
of them form a charge ordered PDW state, with preferred doping of
$3/16$ or $1/4$. This charge ordered PDW state either phase
separate from the SC state, or coexists with it, but either way,
the superfluid density is reduced because of the repulsive
coupling between the two forms of order. Therefore, this picture
predicts a new charge ordered state on the overdoped side, which
should be tested experimentally. Since the magnon density
decreases monotonically with doping, the charge ordered states on
the overdoped side may not be AF ordered, which makes it hard to
observe by neutron scattering. Furthermore, on the overdoped side,
our purely bosonic model also becomes less accurate, and a fully
quantitative theory has to include the low energy fermionic
excitations.

\subsection{The $1/8$ anomaly and pressure experiments}
In the $LSCO$ systems, $T_c$ has a pronounced dip near doping of
$\delta=1/8$\cite{Moodenbaugh1988}. More recently, it has been
demonstrated that the competition between the nearly insulating
$1/8$ phase and the SC phase in $LSCO$ family can be controlled by
pressure. There are two different approaches. One is to apply
hydrostatic or uniaxial pressure on single
crystal\cite{Arumugam2002}. In this way, the pressure in $ab$
plane increases $T_c$ while the pressure in $c$ direction
decreases $T_c$. The other is to introduce compressive or
expansive strain into thin films with the help of the lattice
mismatch between the film and substrate\cite{Sato2000}. Enhanced
$T_c$ and disappearance of the $1/8$ anomaly for compressed film
and strong reduction of $T_c$ for expanded film were
observed\cite{Sato2000}.

Our model reproduces the $\delta=1/8$ effect for $J_c<1$, as one
can see from Fig. \ref{fig:SC}. The pronounced dip is caused from
the competition between the SC state and the insulating PDW states.
Varying doping correspond to the ``B2" or ``B3" traces as depicted
in Fig. \ref{FIG-types}. On the other hand, pressure in the $ab$
plane reduces the lattice constant, thus increases the hopping
term $J_c$. Therefore, applying pressure in the $ab$ plane is
equivalent to following a ``class A2" trace starting from a small
$J_c$ in the global phase diagram. The doping dependence of $T_c$
for different films given in Fig. 3 of Sato {\it et
al}\cite{Sato2000} shares many common features with our doping
dependence of SC order parameter, both MF results (Fig.
\ref{FIG-SC-MF}) and QMC results (Fig. \ref{fig:SC}). The
destructive effect on $T_c$ of the pressure in the $c$-direction
can also be understood in terms of the Poisson effect. The lattice
constant in $ab$ plane will increase when the sample is compressed
in the $c$ direction. This will lead to the decrease of $T_c$ as
argued previously.

Similar to the ``class A2" trace, ``class A3" trace can also be
realized by applying pressure. Therefore, we predict a similar
pressure induced superconductor-to-insulator transition at
$\delta=1/16$.

In this way, the pressure effect on the $1/8$ anomaly is
understood in terms of the bosonic superfluid-to-insulator
transition at the fixed density of $\delta=1/8$. Standard
predictions on the superfluid-to-insulator transition applies to
the $\delta=1/8$ transition. In addition, we argued that the tip
of the $\delta=1/8$ lobe can possibly have the full quantum
$SO(5)$ symmetry. This prediction can be experimentally tested by
comparing both the static and dynamic charge and magnetic
responses, as we have discussed in section II.B.

\subsection{The vortex phase and the ground state above $H_{c2}$}

The ``class A2" and ``class A3" traces can also be approached in
the cuprates by applying a magnetic field along the $c$-axis,
which effectively reduces the hopping term $J_c$. In underdoped
$LSCO$ samples, we predict that the magnetic field destroys SC
order by localizing hole-pairs into a PDW state. This naturally
leads to the field-induced insulating behavior in underdoped
$LSCO$\cite{HAWTHORN2003,SUN2003}. For the $YBCO$ and $BSCO$
systems, the magnetic field could drive the hole-pairs into a
disordered state before the lobe of $\delta=1/8$ or
$\delta=1/16$ is reached.

Recently, a striking feature is revealed in the STM experiments by
Hoffman {\it et al.}\cite{HOFFMAN2002}, where the local density of
states (DOS) near the vortex core show a two dimensional
checkerboard-like modulation with a $4a\times 4a$ charge unit
cell. Here $a$ is the lattice spacing of the $CuO_2$ plane. This
modulation decays exponentially away from the center of the vortex
core, with a decay length of about $10a$. A similar pattern has
also been seen in the absence of the applied magnetic
field\cite{HOWALD2003}, possibly induced by the impurities at the
surface. In the case of vortex core experiment, SC is destroyed by
the magnetic field, and the nature of the competing state is
revealed. The case of impurity scattering is more complex, and the
experimental observation can also be interpreted as due to
quasi-particle interference\cite{WANG2003,MCELROY2003,MCELROY2003A}.

The $\delta=1/8$ insulating PDW state was proposed as an
explanation for the STM measurements\cite{CHEN2002A}. This state
has the $4a\times 4a$ checkerboard symmetry as observed in the
experiment, and the doping level for the insulating $\delta=1/8$
state is reasonably close to the optimal doping level of the
cuprates. On the other hand, if the holes themselves, rather than
the hole pairs, form a Wigner crystal state, the periodicity of
the charge ordering would be larger by a factor of $\sqrt{2}$,
inconsistent with the experiment. Therefore, by forming the Wigner
crystal state of the hole pairs rather than the holes themselves,
the doping level can be compatible with the observed size of the
charge unit cell. In Ref.[\onlinecite{CHEN2002A}], the hole pair
checkerboard state was established by a mean field calculation. A
main result of the present work is to establish the existence of
this state by QMC calculations.

Our calculations as summarized in Fig. \ref{FIG-phase-MF} and
\ref{FIG-Phase-QMC} show that the charge ordered insulating states
are also accompanied by AF magnetic order. Enhanced AF fluctuation
in the vortex state was originally predicted within the $SO(5)$
theory in Ref.[ \onlinecite{ZHANG1997,AROVAS1997}], and has been
experimentally observed in a number of HTSC cuprate families by a
variety of experimental
techniques\cite{KATANO2000,MITROVIC2001,LAKE2001,LAKE2002,
MILLER2002,KHAYKOVICH2002,MITROVIC2003}. More recently, AF order
has been detected by neutron scattering above $H_{c2}$, in
electron doped cuprates\cite{KANG2003,FUJITA2003}. This magnetic
field induced quantum phase transition from the SC state to the AF
state correspond to the ``A2" trace as depicted in Fig.
\ref{FIG-types}.

\subsection{Charge localization and suppression of thermal conductivity}
In previous subsections, we argue that the competition between PDW
and SC can be tuned by applying pressure or magnetic field. In
particular, we show that the PDW ordering of hole-pairs can
develop in the vortex core. The localized hole-pairs in a PDW state
are expected to have little contribution to the transport
properties such as thermal conductivity. This leads to the
argument of the suppression of thermal conductivity by applying a
magnetic field in $c$-axis below some temperature $T_K$. Since the
onset temperature $T_K$ is expected to be proportional to the
superfluid density, $T_K$ has weak dependence on the magnetic
field and follows $T_c$ in the underdoped cuprates. Moreover, the
closer the system is to the charge ordered insulating state such
as the $1/8$ PDW state, the smaller the suppression effect would
be. Finally, the in-plane magnetic field has little effect on the
thermal conductivity due to the fact that it does not create
vortex in the $ab$ plane.

Therefore, trace ``A2" in our global phase diagram and the charge
localization into a hole pair crystal can possibly explain the
recent experiment on the suppression of thermal conductivity by
applying a magnetic field in $c$-direction\cite{KUDO2003}. We also
predict that applying pressure will also induce the suppression or
enhancement of the thermal conductivity around the $1/8$ doping,
assuming the pressure will not induce strong effect of the lattice
structure which can also change the thermal conductivity.

\subsection{1/16 doping}
\begin{figure}
  \includegraphics[scale=1]{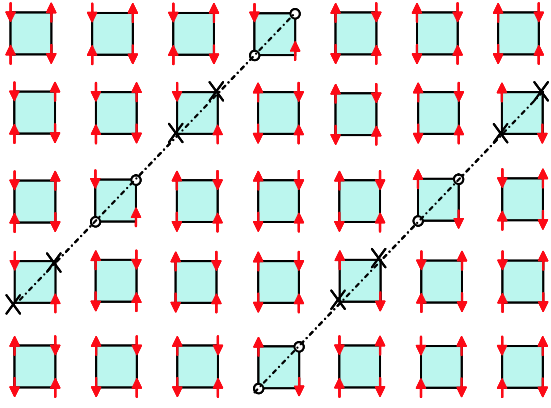}
\caption{The hole-pair checkerboard state with anti-phase magnetic domain at doping
level $\delta=1/16$. Circles denote holes and arrows denote spins
while black crosses denote spins in a singlet bond.}
\label{FIG-diagonal-stripe}
\end{figure}

\begin{figure}
  \includegraphics[scale=1]{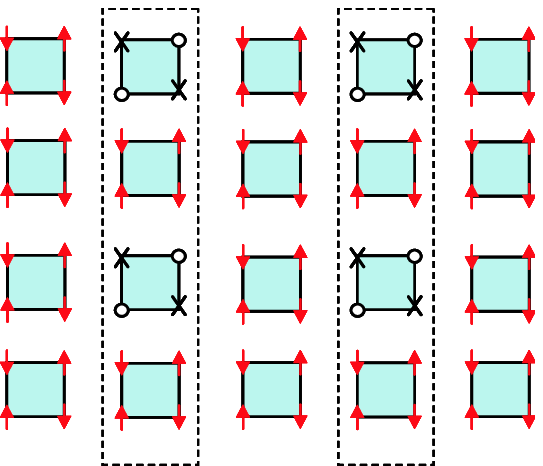}
\caption{The hole-pair checkerboard state with in-phase magnetic
domain at doping level $\delta=1/8$. Circles denote holes and
arrows denote spins while black crosses denote spins in a singlet
bond.}
 \label{FIG-InPhase-stripe}
\end{figure}

As discussed in section \ref{SECTION-heuristic}, additional
insulating lobes at $\delta=1/16$ and $\delta=3/16$ doping levels
are predicted if interactions are more extended than the nnn
interactions included in this work. The charge and spin ordering
patterns for this state are depicted in Fig.
\ref{FIG-diagonal-stripe}.
 Even though the current paper
does not investigate this type of more extended interactions
explicitly due to numerical complexities, it is clear that the
physics of these PDW states are similar to the $1/8$ doping.
Preliminary evidence for the $\delta=1/16$ insulating state exists
for the $LSCO$ material\cite{KIM2001,ZHOU2003}.
 As we see from Fig. \ref{FIG-diagonal-stripe} and \ref{FIG-InPhase-stripe},
the charge ordering pattern rotates from diagonal at $\delta=1/16$
to square at $\delta=1/8$. This transition between states with
different rotational symmetries could possibly be related to the
diagonal to square transition observed in the neutron scattering
experiments at $\delta \sim 5\%$ in the $LSCO$ material\cite{FUJITA2002,MATSUDA2000}.

\begin{figure}
  \includegraphics[scale=1]{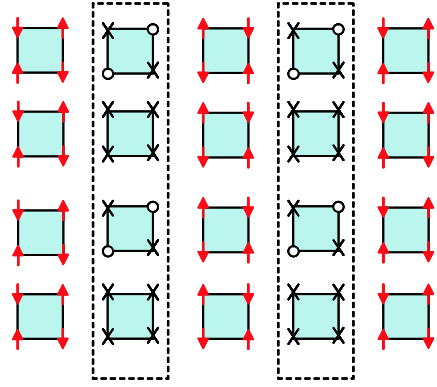}
\caption{The hole-pair checkerboard state with anti-phase magnetic
domain at doping level $\delta=1/8$. Circles denote holes and
arrows denote spins while black crosses denote spins in a singlet
bond.} \label{FIG-magnetic-stripe}
\end{figure}

\subsection{Magnetic order}

In this paper, we focused our discussion on the checkerboard
charge order at $\delta=1/8$. Strictly within our model, the
accompanied AF magnetic order is commensurate, as sketched in Fig.
\ref{FIG-InPhase-stripe}. This type of magnetic structure is
consistent with the recently observed field induced magnetic order
in the $NCCO$ material\cite{KANG2003,FUJITA2003}, but not
consistent with the magnetic structure observed in the $LSCO$
material, which has anti-phase domain walls. We would like to
stress that this is not a limitation on the fundamental approach
taken here. As shown in Fig. \ref{FIG-magnetic-stripe},
checkerboard charge order of the hole pairs can be fully
consistent with the AF magnetic structure with anti-phase domain
walls. However, stability of this type of magnetic structure
requires more extended magnetic interactions. Since the complexity
of both the CORE algorithm and the QMC increases substantially, we
have not yet been able to derive such extended interactions from
the microscopic models, and simulate them with QMC.

\subsection{Coexistence phases}
While the simple $SO(5)$ model\cite{ZHANG1997} predicts the
coexistence phases of AF and SC, more regions with coexisting
charge, AF and SC orders are predicted in the global phase diagram
of Fig. \ref{FIG-phase-MF} and Fig. \ref{FIG-Phase-QMC}. In
particular, the ``type 1" first order transition from undoped AF
state to SC state predicted by simple $SO(5)$ model is turned into
two ``type 2" second order transitions by the interactions $V_c$
and $V_c'$, which is consistent with previous
study\cite{DORNEICH2002}. While the width of these coexistence
regions is model dependent, coexistence phases are important
qualitative predictions of our theory. Experiments on cuprates
have indeed suggested such coexistence
phases\cite{BREWER1989,NIEDERMAYER1998,SONIER2001,SIDIS2001,MOOK2002,Miller2003}.

As we see from Fig. \ref{FIG-AF-MF}, the AF order disappears
around $\delta=0.10$. This is consistent with the value obtained
from the $t$-$J$ model\cite{HIMEDA1999}. The width of the AF/SC
coexistence phase largely depends on the values of $J_\pi$ and
$V_\pi$. As determined by CORE method, $J_\pi$ is negative and
$V_\pi$ is positive. Thus, at the mean-field level, both $J_\pi$
and $V_\pi$ terms induce a repulsion between AF and SC in the
mixed state of coexisting AF and SC. When these terms are
included, one would expect a smaller region of the AF/SC
coexistence state of in the global phase diagram. On the other
hand, these two terms have different effects on the checkerboard
state. Since the mean-field value of $\pi$-operator in the PDW
state is zero, the $J_\pi$ interaction will not induce any
interaction terms at the mean-field level. In contrast, the
$V_\pi$ term effectively changes the local chemical potential of
hole-pairs (magnons) due to the nonzero mean-field value of
magnons (hole-pairs) on nn plaquette. Therefore, $V_\pi$ also
reduces the height of the insulating PDW lobe.

\section{Conclusions}\label{SECTION-Conclusion}
Starting from commonly used microscopic models of high $T_c$
cuprates, an effective bosonic model can be derived by the CORE
algorithm\cite{ALTMAN2002A,CAPPONI2002}. In addition to the simple
interactions included in the original projected $SO(5)$
model\cite{ZHANG1999,DORNEICH2002,JOSTINGMEIER2002}, extended
interactions play an important role in determining the global
phase diagram of the model. This model can be studied
systematically by the analytical mean field theory and by the QMC
method, thanks to the absence of the minus sign problem. The
global phase diagram consists of self-similar insulating PDW phases
at rational filling fractions, immersed in the background of the
uniform SC phase, as depicted schematically in Fig.
\ref{FIG-types}. Different families of cuprates are attributed to
different traces in the global phase diagram. The overall topology
of the phase diagram obtained from our model agrees well with the
experiments, and so are the behaviors of various physical
quantities. Inclusion of longer ranged interactions could bring
detailed and quantitative agreements with the cuprate phase
diagram.

\section{acknowledgments}
We would like to acknowledge useful discussions with Drs. A.
Auerbach, E. Demler, W. Hanke, J.P. Hu, S. Kivelson, E. Mukamel,
G. Schmid, M. Troyer and C.J. Wu. This work is supported by the NSF under
grant numbers DMR-9814289, and the US Department of Energy, Office of Basic
Energy Sciences under contract DE-AC03-76SF00515 and by the Swiss
National Science Foundation. HDC is also supported by a Stanford
Graduate Fellowship.
Part of the simulations were performed on the Asgard beowulf cluster
at ETH Z\"urich and at SLAC.

%\bibliography{global,extra}

\end{document}